\documentclass[11pt]{article}
\usepackage{amsmath, amssymb}
\usepackage{slashed}
\usepackage{verbatim}
\usepackage{latexsym}
\usepackage{euscript}
\usepackage{amsfonts}
\usepackage{verbatim}
\usepackage[]{array}
\usepackage[]{mathrsfs}
\usepackage[]{tensor}
\usepackage[symbol]{footmisc}

\def\ket#1{|#1 \rangle}
\def\be{\begin{eqnarray}}
\def\ee{\end{eqnarray}}
\def\0{\nonumber}
\def\del{\partial}

\def\ve{\varepsilon}

\newcommand\EL{\EuScript{L}}
\newcommand\EF{\EuScript{F}}
\begin{document}

\begin{flushright}
SISSA/63/2014/FISI\\
hep-th/1412.xxxx
\end{flushright}
\vskip 2cm

\begin{center}

{\LARGE {\bf  Comments on lump solutions in SFT}}

\vskip 1cm

{\large Loriano Bonora and Driba D.Tolla}

{}~\\

\quad \\

{\em ~ International School for Advanced Studies (SISSA),}\\

{\em  Via Bonomea 265, 34136 Trieste, Italy and INFN, Sezione di Trieste}\\

{\tt bonora@sissa.it, dribatolla@gmail.com}

 \end{center}

 \vskip 2cm {\bf Abstract.}
We analyze a recently proposed scheme to construct analytic lump solutions in open SFT.
We argue that in order for the scheme to be operative and guarantee background
independence it must be implemented in the same 2D conformal
field theory in which SFT is formulated. We outline and discuss two different possible
approaches. Next we reconsider an older proposal for analytic lump solutions and
implement a few improvements. In the course of the analysis we formulate a
distinction between regular and
singular gauge transformations and advocate the necessity of defining a topology
in the space of string fields.

\eject

\section{Introduction}

After the discovery \cite{Schnabl05,Ellwood}  of the first analytic solution of
open SFT \`a la Witten, \cite{W},
which links the perturbative vacuum to the tachyon vacuum,
there has been a considerable number of papers devoted to related solutions,
\cite{Okawa:2006vm,ES}
and to marginal deformations thereof,
\cite{KORZ,Okawa3,Schnabl:2007az,Fuchs3,Lee:2007ns,Kiermaier:2007vu,KOS,
Maccaferri:2014cpa}. The literature concerning
analytic lump solutions, i.e. analytic
solutions interpretable as lower dimensional branes (meant to complete the
analytic proof of the three conjectures by Sen, \cite{Sen}),
 is instead poorer. There have
been essentially two
attempts to find such analytic solutions: the first is the so-called BMT
proposal, \cite{BMT} and
\cite{BGT1,BGT2}, the second is the most recent one formulated in \cite{EM},
which we will refer to as EM.
They are both modelled on the Erler-Schnabl (ES) solution,
\cite{ES}, an alternative simpler
formulation of the original tachyon vacuum (TV) solution (for recent reviews on
the whole subject, see
\cite{Fuchs,Schnabl,Okawa,Bonora}). In \cite{EM} the construction is based
on previous results on correlators involving boundary condition changing  (bcc)
operators
and on a set of (implicit) prescriptions laid down in order for the solution to
satisfy the SFT equation of motion. As we shall see, hidden behind this is the
risk of
background dependence. The only way to clarify this issue is to implement the EM
approach
in an explicit 2D field theory formulation similar to the one in
\cite{ES},\cite{KOS}
and \cite{BMT}, consistent with the original one in which the SFT is formulated.
The first part
of our paper is an attempt in this direction. We clarify the problems behind the
EM
proposal and try to solve them. We do not succeed in carrying out this task,
but, nevertheless,
we think our scrutiny may be instructive. We then turn to the BMT proposal and,
in the second
part of the paper, we discuss some aspects of the latter and present a few
improvements. In the course of the
analysis we propose a definite distinction between singular and regular BRST
transformations,
which turns out to be instrumental in clarifying some confused issues present in
analytic
solutions of SFT.

The paper is organized as follows. Sections 2,3 and 6 are short introductions to
the ES solution and to the EM and BMT proposals. Section 4 is devoted to a
detailed
presentation of the bcc operators used in the EM proposal. Section 5 contains
our (failed)
attempt to implement the  EM approach in an explicit field (oscillator)
formulation.
Section 7 contains a digression on gauge transformations and identity based
solutions, which is needed for the
subsequent developments in section 8, where they are applied to the BMT
proposal.
Section 9 contains some conclusions.

\section{Short review of the Erler--Schnabl solution}

Even though the content of this section is very well-known, we nevertheless
briefly review the classical solution and its properties found by Erler and
Schnabl \cite{ES}, given its paradigmatic role with respect to the proposed lump solutions.
To start with let us introduce the $\{K, B, c\}$ algebra, where
\begin{equation} \label{1}
K= \frac {\pi}2 K_1^L|I\rangle,\quad\quad B=\frac {\pi}2 B_1^L|I\rangle,
\quad\quad c= c(1) |I\rangle  \;,
\end{equation}
where the argument of the $c$ ghost field refers to the arctan frame and
\be
K_1^L =\frac12 K_1+\frac{1}{\pi} ({\cal L}_0+{\cal L}^\dagger_0) \;,\quad\quad
B_1^L = \frac12 B_1+\frac{1}{\pi} ({\cal B}_0+{\cal B}^\dagger_0)
\;,\label{K,B,c}
\ee
and the $K_1, {\cal L}_0$ and $B_1, {\cal B}_0$ are expressed in terms of the
worldsheet energy--momentum tensor and the $b$ ghost field in the following way
\be
{\cal L}_0 & =& L_0+\sum_{k=1}^{\infty} \frac{2 (-1)^{k+1}}{4k^2-1}L_{2k} \; ,
{\cal B}_0  = b_0+\sum_{k=1}^{\infty} \frac{2 (-1)^{k+1}}{4k^2-1}b_{2k} \; ,\0\\
K_1 & =& L_1 + L_{-1} \; ,\quad
B_1  = b_1 + b_{-1} \;.\0
\ee
Then the string fields $K$, $B$ and $c$ generate an algebra with the following
commutation relations and BRST variations
\be\label{KBc1}
&&[K,B]=0,\quad\quad \{B,c\}+ =1, \quad\quad [K,c]= \partial c,\quad\quad
\{B,\partial c\}+=0 ,\0\\
 &&B^2=0=c^2,\quad\quad
 Q\,B=K,\quad\quad QK=0,\quad\quad Q\, c=cKc=c\del c\label{KBc}
\ee
In the above the juxtaposition of any two symbols represents the star product,
whose
symbol has been understood. The commutators are taken with respect to the star
product.

The Erler--Schnabl (ES) solution is given by
\begin{equation}
\psi_{ES} = c B(K+1)c \frac 1{K+1}
\end{equation}
The solution can be formally written via a {\it singular} gauge transformation
of the perturbative vacuum (see \cite{Okawa:2006vm,Erler:2012qn})
\begin{equation}
\psi_{ES} \ = \ U_0 Q U_0^{-1} \;,
\end{equation}
with
\begin{equation}
U_0=1-\frac{1}{K+1}Bc \;\; \text{and} \;\; U_0^{-1} = 1+\frac{1}{K}Bc \;.
\end{equation}

In order to show that the ES solution satisfies Sen's first conjecture one
computes the energy corresponding to it. The energy density is (in the sequel we
set  $g_o=1$)
\be \nonumber
E[\psi_{ES}]&=\frac 16 \langle \psi_0, Q\psi_0\rangle= \frac 16 \langle (c+cKBc)
\frac 1{K+1}\,cKc\,\frac 1{K+1}\rangle \\
&= \frac 16 \langle c \,\frac 1{K+1}\,cKc\,\frac 1{K+1}\rangle - \langle
Q\left(Bc \frac 1{K+1}\,cKc\,\frac 1{K+1}\right)\rangle \;.
\ee
The last term vanishes since it is BRST exact and the first term in the second
line gives
\be \nonumber
E[\psi_{ES}]& =\frac 16 \int_0^\infty dt_1dt_2 \, e^{-t_1-t_2}\langle c\,
e^{-t_1 K}\,c \partial c \,e^{-t_2K}\rangle_{C_{t_1+t_2}} \\ \nonumber
& =-\frac 16 \int_0^\infty dt_1 dt_2 \, e^{-t_1-t_2} \, \frac
{(t_1+t_2)^2}{\pi^2} \sin^2\left(\frac {\pi t_1}{t_1+t_2}\right)\\
\label{Epsi01}
& =-\frac 1{2\pi^2}
\ee
where $C_t$ denotes a cylinder in the $\arctan$ frame of circumference $t$, and
we have used the Schwinger parametrization
\begin{equation}
\frac{1}{1+K} = \int_{0}^{\infty} e^{-t(1+K)} \;.\0
\end{equation}
In passing from the first line to the second line of (\ref{Epsi01}) one starts
from the correlator $\langle c(z_1) c\partial c(z_2)\rangle =-(z_1-z_2)^2 \0$
in the upper half--plane, and maps it to the arctan frame via the map $\xi=
\arctan(z)$, so that it becomes
\begin{equation}
\langle c(\xi_1) c\partial c(\xi_2)\rangle =-\sin^2(\xi_1-\xi_2)\0
\end{equation}
in the cylinder $C_\pi$. Finally one rescales $\xi \to \frac {\ell}{\pi} \xi $
in order to map to a cylinder $C_\ell$.

Equation (\ref{Epsi01}) means that the ES solution correctly reproduces
minus the $D25$--brane tension, thus identifies the tachyon condensation vacuum.
The Erler--Schnabl solution does not support open string states, i.e. this
solution agrees with Sen's second conjecture. A simple way to show that a given
solution $\psi_0$ does not support open string states is to find a homotopy
operator ${\cal A}$ such that
\begin{equation}
{\cal Q}_{\psi_0}{\cal A}=1 \;,\label{homop}
\end{equation}
with
\begin{equation}
{\cal Q}_{\psi_0} {\cal A}=Q{\cal A} +\psi_0{\cal A} +{\cal A} \psi_0 \;.
\end{equation}
for, if ${\cal Q}_{\psi_0} \phi=0$ this implies that $\phi= {\cal Q}_{\psi_0}
\chi$  with $\chi= {\cal A}\phi$.
For the ES solution, the homotopy operator exists and has the form
\cite{Ellwood}
\begin{equation}
{\cal A}=B \frac{1}{1+K}.
\end{equation}

\section{The EM proposal}

Let us briefly review here the recent EM proposal, at least for the (essential)
aspects we wish to discuss.
The EM proposal is constructed by introducing in the ES solution
boundary condition changing  (bcc)
operators, in analogy to the Kiermaier, Okawa, and Soler (KOS) solution,
for marginal deformations (see  \cite{KOS} and below).
The KOS solution is written in terms of
boundary condition changing  (bcc) operators $\sigma$ and $\bar\sigma$
satisfying
the following OPE
\begin{align}\label{OPE1}
\sigma\bar\sigma=\bar\sigma\sigma=1.
\end{align}
In EM the relevant bcc operators are not the same as in KOS, of course,
rather they change the string boundary conditions from N to D and viceversa.
The property (\ref{OPE1}) is crucial for the solution to satisfy the equation of
motion and, at first,
it seems difficult to extend the procedure to solutions which describe BCFTs
that are not connected to the original BCFT by a marginal deformation. The
reason is that, generic bcc operators relating two BCFTs have non trivial OPE
and cannot satisfy \eqref{OPE1}. In order to overcome this difficulty the
authors of \cite{EM} tensor those bcc operators with plane-wave factors and
introduce
modified bcc operators as follows
 \begin{align}
 \sigma(s)=\sigma_\ast(s) e^{ i\sqrt{h}X^0(s)},\qquad
\bar\sigma(s)=\bar\sigma_\ast(s) e^{- i\sqrt{h}X^0(s)}.
\end{align}
where $\sigma_\ast(s)$ and $\bar\sigma_\ast(s) $ are the bcc operators relating
the two BCFTs and they are primary operators of conformal dimension $h$. In
particular, for the lump solution they are given by
$\sigma_\ast=\sigma_{\!N\!D},~\bar\sigma_\ast=\sigma_{\!D\!N}$ and
$h=\frac1{16}$. The
modified bcc operators satisfy

\begin{align}\label{OPE2}
\sigma\bar\sigma={\rm finite}~,\quad\bar\sigma\sigma=1.
\end{align}
EM assumes that the BRST variations of the modified bcc operators are
\begin{align}\label{BRST1}
Q\sigma=c\partial\sigma=c[K,\sigma],\qquad Q\bar\sigma=c\partial\bar\sigma=c[K,\bar\sigma]
\end{align}
where in \cite{EM} $Q$ is the BRST operator of the original D25-brane BCFT.
This is
one of the critical aspects we must discuss. In fact, as we shall see,
there is no a priori guarantee that the action of $Q$ is well-defined on
$\sigma, \bar\sigma$.

Continuing the presentation of the EM proposal, the EM solution for the equation of motion at
the tachyon vacuum is given by
 \begin{align}
\Phi_0=-\Sigma\Psi_0\bar\Sigma\label{Phi0}
\end{align}
where
\begin{align}
&\Psi_0=\frac1{\sqrt{1+K }}c(1+K )Bc\frac1{\sqrt{1+K }},\0\\
&\Sigma={\cal Q} \Big(\frac1{\sqrt{1+K }}B\sigma\frac1{\sqrt{1+K }}\Big),\quad
\bar\Sigma={\cal
Q} \Big(\frac1{\sqrt{1+K }}B\bar\sigma\frac1{\sqrt{1+K }}\Big).\label{
Psi0}
\end{align}
Here ${\cal Q} =Q +[\Psi_0,\,\,]$  is the BRST operators at the tachyon vacuum. Explicitly,
the proposed solution is given by
\begin{align}\label{Phi0exp}
\Phi_0=-\frac1{\sqrt{1+K }}c(1+K )\sigma \frac B{1+K }\bar\sigma (1+K )c\frac1{\sqrt{1+K}}
\end{align}

The solution to the equation of motion at the perturbative vacuum is
\begin{align}
&\Psi=\Psi_0+\Phi_0\label{sol}
\end{align}
and its energy is given by
\begin{align}
E=-S[\Psi]=-\frac{g_0}{2\pi^2}+\frac16{\rm Tr}[\Phi_0^3]
=-\frac{g_0}{2\pi^2}-\frac16{\rm
Tr}[(\Psi_0)^3]=-\frac{g_0}{2\pi^2}+\frac{g_\ast}{2\pi^2}.\label{Phienergy}
\end{align}
Using previous results in the literature, in particular \cite{GNS}, the authors
of \cite{EM}
were able to show that $g_\ast-g_0$ is consistent with the difference between
the tension of a D25 and a D24 brane.

In the EM ansatz many details are understood, and we must ask:
{\it Is it possible to implement this formulation in a concrete field theory
formalism, consistent with that of SFT?} As it turns out
this formulation is deceptively simple.
In the following we would like to bring to light the hidden details and assess
their validity. To start with, two BCFT's are mentioned in
\cite{EM}, BCFT$_0$ and BCFT$_\ast$, and a rule is declared according to which,
when writing the solution, any operator that appears to the left of
$\sigma$ or to the right of $\bar\sigma$ belongs to BCFT$_0$, while operators
which appears to the right of $\sigma$ or to the left of $\bar\sigma$ belong to
BCFT$_\ast$, but no distinction is made to specify what string fields belong to
the former and what to the latter. It may well be that the $Q$ and $K$ that are defined on
BCFT$_0$ are the same that live on BCFT$_\ast$ (this is one of the hypotheses we
will consider later on). But this is a crucial aspects that must be
carefully justified. As we will see more clearly later on, the BCFT$_0$
is the BCFT with NN boundary conditions for all directions, while BCFT$_\ast$
is characterized by DD boundary conditions along one space direction, the 25-th,
say, and NN along the remaining ones.
To keep track of this we will simply label $Q$ and $K$ in these theories
with NN and DD, respectively.

Thus, in order to be consistent with the EM rules, when passing through
$\sigma$, operators like
$Q_{\rm NN}$ and $K_{\rm NN}$, become $Q_{\rm DD}$ and $K_{\rm DD}$, and they
switch back to $Q_{\rm NN}$ and $K_{\rm NN}$ when they
pass through $\bar\sigma$. We will discuss later on how it is possible to
make sense of such switching of operators. In the rest of this section we would like to
make it evident that considering only two BCFT's, BCFT$_0$ and BCFT$_\ast$, is
not enough.

In a first attempt to write the EM proposal (\ref{Phi0}) in a more precise way, we
rewrite it as
 \begin{align}
\Phi_0=-\Sigma\Psi_0^\ast\bar\Sigma\label{Phi*}
\end{align}
where
\begin{align}
&\Psi_0^\ast=\frac1{\sqrt{1+K_{\rm DD}}}c(1+K_{\rm DD})Bc\frac1{\sqrt{1+K_{\rm
DD}}},\0\\
&\Sigma={\cal Q}_{\rm NN}\Big(\frac1{\sqrt{1+K_{\rm
NN}}}B\sigma\frac1{\sqrt{1+K_{\rm DD}}}\Big),\quad
\bar\Sigma={\cal
Q}_{\rm DD}\Big(\frac1{\sqrt{1+K_{\rm DD}}}B\bar\sigma\frac1{\sqrt{1+K_{\rm
NN}}}\Big).\label{Psi0*}
\end{align}
 with ${\cal Q}_{\rm NN}=Q_{\rm NN}+[\Psi_0,]$ and ${\cal Q}_{\rm DD}=Q_{\rm
DD}+[\Psi_0^\ast,]$ are the BRST operators at the tachyon vacuum. Explicitly,
the ansatz is given by
 \begin{align}\label{Phi0*}
 \Phi_0=-\frac1{\sqrt{1+K_{\rm NN}}}c(1+K_{\rm NN})\sigma \frac B{1+K_{\rm
DD}}\bar\sigma (1+K_{\rm NN})c\frac1{\sqrt{1+K_{\rm NN}}}
 \end{align}

The solution to the equation of motion at the perturbative vacuum is
\begin{align}
&\Psi=\Psi_0+\Phi_0\label{sol*}
\end{align}
where
$ \Psi_0=\frac1{\sqrt{1+K_{\rm NN}}}c(1+K_{\rm NN})Bc\frac1{\sqrt{1+K_{\rm
NN}}}$ is still the ES TV solution,
and its energy is expected to yield
\begin{align}
E=-S[\Psi]&=-\frac{g_0}{2\pi^2}+\frac16{\rm Tr}[\Phi_0^3]\0\\
&=-\frac{g_0}{2\pi^2}-\frac16{\rm
Tr}[(\Psi_0^{\ast})^3]=-\frac{g_0}{2\pi^2}+\frac{g_\ast}{2\pi^2}.
\label{Phienergy*}
\end{align}

Now that we have heuristically rewritten the EM rules in a more explicit form
let us  point out some difficulties. The fact that the EM ansatz yields
the expected result depends
crucially on enlarging the $K,B,c$ algebra by including $\sigma$ and $\bar
\sigma$. The $K, B, c$ algebras (for zero momentum states)
for the NN and DD cases are identical (see below). For clarity let us write them once more
\begin{align}\label{KBc2}
&\{B,c\}=0, \quad \{B,\partial
c\}=0,\quad [K_{ii},B]=0,\quad [K_{ii},c]=\partial c,
\end{align}
where $ii$ denotes either NN or DD. The action of the BRST operators on those
quantities should mimic the equation
\eqref{KBc1}
\begin{align}\label{BRST3}
&\quad Q_{ii} B=K_{ii},\quad Q_{ii} K_{ii}=0,\quad Q_{ii} c=cK_{ii} c.
\end{align}
From equation \eqref{1} we directly get the representation of two
of the
elements of the algebra \eqref{KBc2}, but the representation of $K_{\rm DD}$
should be obtained as the BRST variation of $B$:
 \begin{align}\label{BRST4}
\frac\pi2 Q_{\rm DD}
B_1^L|I\rangle&=\frac\pi2\{Q_{\rm DD},B_1^L\}|I\rangle-\frac\pi2 B_1^LQ_{\rm DD}
|I\rangle
=\frac\pi2K_{1,\rm DD}^{L}|I\rangle-\frac\pi2 B_1^LQ_{\rm DD} |I\rangle
\end{align}
So we see that the BRST operator $Q_{\rm DD}$
must annihilate the identity state $|I\rangle$ or any wedge state in
general, otherwise the definition of $K_{\rm DD}$ as a BRST variation of $B$ can
not be realized at the representation level. Is this the case? The answer is not obvious.

Another concerning issue is related to the computation of
physical observables for the solution $\Psi_0^\ast$. In order to calculate the
energy or the closed string overlap, one needs to use the following Schwinger
parametrization for
the inverse of $1+K_{\rm DD}$:
\begin{align}
\frac1{1+K_{\rm DD}}=\int_0^\infty dt e^{-t}e^{-tK_{\rm DD}}.
\end{align}
In the ES solution, we replace $e^{-tK}$ by $\Omega^t$,
where $\Omega$ is the SL$(2,{\mathbb R})$ invariant vacuum, which defines the
wedge states. Now, in the presence of the bcc operators, is the
vacuum SL$(2,{\mathbb R})$ invariant and does the new vacuum
have the right properties to define new wedge states?

A third problem arises when we try to verify the equation of motion for \eqref{Phi0*}.
In our notation the equation motion at the TV is
\begin{align}\label{TVEoM}
{\cal Q}_{\rm NN}\Phi+\Phi^2=0\Rightarrow {Q}_{\rm
NN}\Phi+\{\Psi_0,\Phi\}+\Phi^2=0.
\end{align}
According to the EM rules rewritten above, the BRST operator acts on $\Phi_0$ as follows
\begin{align}\label{QPhi0}
 Q_{\rm NN}\Phi_0&=-\frac1{\sqrt{1+K_{\rm NN}}}(Q_{\rm NN}c)(1+K_{\rm NN})\sigma
\frac B{1+K_{\rm DD}}\bar\sigma (1+K_{\rm NN})c\frac1{\sqrt{1+K_{\rm NN}}}\0\\
 &+\frac1{\sqrt{1+K_{\rm NN}}}c(1+K_{\rm NN})(Q\,\sigma)
\frac B{1+K_{\rm DD}}\bar\sigma (1+K_{\rm NN})c\frac1{\sqrt{1+K_{\rm NN}}}\0\\
 &+\frac1{\sqrt{1+K_{\rm NN}}}c(1+K_{\rm NN})\sigma \frac {(Q_{\rm DD}B)}
{1+K_{\rm DD}}\bar\sigma (1+K_{\rm NN})c\frac1{\sqrt{1+K_{\rm NN}}}\0\\
 &-\frac1{\sqrt{1+K_{\rm NN}}}c(1+K_{\rm NN})\sigma \frac B{1+K_{\rm DD}}(Q\,\bar\sigma)
(1+K_{\rm NN})c\frac1{\sqrt{1+K_{\rm NN}}}\0\\
 &-\frac1{\sqrt{1+K_{\rm NN}}}c(1+K_{\rm NN})\sigma \frac B{1+K_{\rm
DD}}\bar\sigma (1+K_{\rm NN})(Q_{\rm NN}c)\frac1{\sqrt{1+K_{\rm NN}}},
\end{align}
where $Q\,\sigma$, $Q\,\bar \sigma$ is only indicated, but not specified. In fact, in these two cases
$Q$ cannot be neither $Q_{\rm NN}$, nor $Q_{\rm DD}$.
In fact we will see in the next section that
$\sigma$ and $\bar\sigma$ belong to two additional BCFT's: the
BCFT with ND boundary condition along one space direction, and the BCFT with DN
boundary condition along one space direction (and NN along all the others). The corresponding
$Q$ and $K$ will be called $Q_{\rm ND},~K_{\rm ND}$ and $Q_{\rm DN},~K_{\rm DN}$,
respectively.

With these new entries it seems to be more appropriate to rewrite \eqref{BRST1} as
\begin{align}\label{BRST2}
Q_{\rm ND}\sigma=c\partial\sigma=c(K_{\rm ND}\sigma-\sigma K_{\rm ND}),\qquad
Q_{\rm DN}\bar\sigma=c\partial\bar\sigma=c(K_{\rm DN}\bar\sigma-\bar\sigma
K_{\rm DN}),
\end{align}
Assuming these BRST variations of $\sigma$ and $\bar\sigma$ and that the star product
is consistent with these rules, we
notice that (\ref{QPhi0}) contains $K_{\rm ND}$ and $K_{\rm DN}$, whereas the other
two terms in \eqref{TVEoM} do not contain such operators. Therefore, the
cancellation among those terms is not possible and the equation of motion
would not be satisfied.

It is clear that the previous minimalistic cosmetic of the EM rules is too simplistic and
only complicates things. A deeper interpretation is necessary. But it is also clear that
if the above issues (among others, see below) are not clarified within a concrete
field theory formalism, the EM ansatz remains abstract. Our aim in the sequel is to interpret
the EM rules in a 2D conformal field theory context, consistent with
the formulation of SFT.

\section{The bcc operator}
\label{sec:bcc}

One crucial ingredient in the EM ansatz are the bcc operators. We devote this section
to a rather detailed description of this subject.

The issue of bcc operators was introduced by Cardy, see \cite{Cardy}, and
subsequently studied and applied by many authors, see in particular
\cite{Zamolo,Hashimoto,GNS,Frohlich}. It is generally believed that
the original and twisted theory are characterized by Hilbert spaces that can be
related
to each other. This is in a sense trivial, because all countable Hilbert spaces
are isomorphic as vector spaces. However a CFT is not simply characterized by
a Hilbert space, but also by the central charge, its primary operators and by
the field theory axioms (locality, for one) and its symmetries.
Therefore a direct connection between two such theories in the form of an
intertwining operator between the two Hilbert spaces (see below) or, even more,
an identification of the two, is far from guaranteed and, if it is
possible,
it is far from trivial to be determined.
This is to stress that it is necessary to analyse in depth the concept and
application of bcc operator,
a type of analysis the existing literature does not abound with.

In the sequel, for definiteness, we will use an explicit
formulation of bcc operator, following in particular \cite{Zamolo}.

The ordinary one-dimensional {NN} string in the complex $z$ plane is ($\alpha'=\frac 12$)
\be
X_{\rm NN}(z,\bar z)= x-\frac i2 \alpha_0 (\ln z+\ln\bar z)+\frac i2 \sum_{n\neq 0} \frac
{\alpha_n}{ n}\big(z^{-n}+\bar z^{-n}\big)\label{XN}
\ee
with $\alpha_0=p$, and $[\alpha_m,\alpha_n]=m\delta_{n+m}$. The relevant holomorphic
propagator is
\be
\langle i\del X_{\rm NN}(z)\, i\del X_{\rm NN}(w) \rangle \sim\frac 1{(z-w)^2}\label{propNN}
\ee
Then one defines the usual $L_n= \frac 12 \sum_{k} :\alpha_{n-k}\alpha_k:$.
The corresponding Virasoro algebra has central charge 1. The
vacuum is defined by $\alpha_n|0\rangle=0$ for $n\geq 0$.

On the other hand the one-dimensional {DD} string is specified by
\be
X_{\rm DD}(z,\bar z)= x_0+\frac i{2\pi} \Delta x\, (\ln z-\ln\bar z)-\frac i2 \sum_{n\neq
0} \frac {\alpha_n}{ n}\big(z^{-n}-\bar z^{-n}\big)\label{XD}
\ee
where $\Delta x=(x_\pi-x_0)$ is the separation between the two D-branes to which
the string is attached. For a single D-brane $\Delta x=0$. The other
$\alpha_n,~n\ne0$  satisfy the same
algebra as the NN string. The holomorphic $U(1)$ current is
\begin{align}
i\partial X(z,\bar z)=-\sum_{n} {\alpha_n}z^{-n-1}, \label{DN}
\end{align}
The relevant propagator and the corresponding Virasoro generators have the same
form as the NN string.
The only difference between the NN and DD cases is the presence of the zero
mode in the former\footnote{For the oscillators we use the same
symbols in the NN and DD case and in the ND and DN case. The reader is invited to keep in mind the
difference.}.
Therefore, the Virasoro algebra is the same for both cases. The BRST operator is
also the same for zero momentum states.

Now we consider the Neumann-Dirichlet (ND) string:
\be
X_{\rm ND}(z,\bar z)= x_0 +\frac i2 \sum_{r\epsilon~ {\mathbb Z}+\frac12} \frac
{\alpha_{r}}{r} \big(z^{-r}+\bar z^{-r}\big)\label{XND}
\ee
and the DN one:
\be
X_{\rm DN}(z,\bar z)= x_0 +\frac i2 \sum_{r\epsilon~ {\mathbb Z}+\frac12} \frac
{\alpha_{r}}{r} \big(z^{-r}-\bar z^{-r}\big)\label{XDN}
\ee
For later use let us define, according to \cite{Zamolo}, the (holomorphic) U(1) current
\be
J(z)= i\del X(z) = \sum_{r\epsilon~ {\mathbb Z}+\frac12} {\alpha_{r}}
z^{-r-1}\label{JND}
\ee
From now on, to avoid misunderstandings, we will replace $\alpha_r$ with the
symbol $J_r$.

Assume the canonical commutator
\be
[J_{r},J_{s}]= r \delta_{r+s}\label{ccND}
\ee
Define the vacuum $|\sigma_\ast\rangle$ by
\be
J_{r}|\sigma_\ast\rangle = 0,\quad\quad{\rm for}\quad
r>0\label{NDvacuum}
\ee
with corresponding definition of normal ordering. The propagator is
\be
\langle J(z) J(w) \rangle \sim \frac 1{(z-w)^2} + {\rm
Reg}\label{propND}
\ee
The Virasoro generators turn out to be
\be
{\EL}_n&=& \frac 12 \sum_{r\epsilon~ {\mathbb Z}+\frac12}
:J_{r}J_{n-r}:\label{LnND}\\
{\EL}_0&=& a_0 + \frac12\sum_{r\epsilon~ {\mathbb Z}+\frac12}
J_{r}J_{-r}\label{L0ND}
\ee
Then we find
\be
[\EL_m, \EL_n] = (m-n)\EL_{n+m} + \frac 1{12} (n^3-n) \label{VirND}
\ee
provided $a_0=\frac 1{16}$. It follows that
\be
{\cal L}_0|\sigma_\ast\rangle= \frac 1{16} |\sigma_\ast\rangle\label{sigmaweight}
\ee

As usual we conclude that in the $c=1$ CFT there is a primary field
$\sigma_\ast(z)$
of weight $\frac 1{16}$. This is the bcc operator.
The first excited state above $\sigma_\ast(z)$ is
\be
|\tau\rangle = J_{-\frac 12}|\sigma_\ast\rangle \label{tau}
\ee
In terms of the current $J$ we have, for small $z$
\be
&&J(z)|\sigma_\ast\rangle =z^{-\frac 12} |\tau\rangle + {\cal O}(z^{\frac
12})\0\\
&&J(z)|\tau\rangle= z^{-\frac 32}|\sigma_\ast\rangle + z^{-\frac 12} J_{-\frac
12}^2|\sigma_\ast\rangle + {\cal O}(z^{\frac 12})\label{OPEsigmatau}
\ee
It is easy to see that $J_{-\frac
12}^2|\sigma_\ast\rangle=2{\cal L}_{-1}|\sigma_\ast\rangle=2
|\sigma'_\ast\rangle $, and the weight of $\tau$ is $\frac 9{16}$.
There is an infinite towers of such states. The spectrum of these states is
completely different from the spectrum of the NN string.

It is clear that we can repeat word by word the same things for the DN string,
whose oscillators
are also half-integral-mode. The vacuum in this case will be denoted by
$|\bar\sigma_\ast\rangle$.

In conclusion in the $c=1$ CFT there is room for four different Hilbert spaces
${\cal H}_{\rm NN}$,
${\cal H}_{\rm DD}$, ${\cal H}_{\rm ND}$ and ${\cal H}_{\rm DN}$. The last two are the same.
The first two
are also identifiable except for the zero mode $\alpha_0$ in the NN case. If
the coordinate $X$ is
compactified on a circle, the momentum is discrete; dually, in the DD case, we
have wrapping modes.
In other words the relevant Hilbert spaces are organized in discrete sectors.
For the moment let us ignore, for simplicity,
such discrete sectors and the momentum in ${\cal H}_{\rm NN}$.
We see that, as far as the EM proposal is concerned we have two kinds of Hilbert
spaces, one built out of
the integral-mode matter oscillators $\alpha_n$ in the 25-th direction, the
other with half-integral-mode oscillators $J_{r}$, beside all the other matter
and ghost oscillators.

At this point, however, it is worth recalling that OSFT is formulated in terms
of NN strings,
that is on the background of the D25-brane. The background independence of OSFT
therefore does not rely on the original formulation, but
on the fact that, starting from it, we can derive all the other possible
backgrounds as
analytic solutions of its equation of motion. This is the OSFT's bet.
It goes without saying that any background independent solution must be
formulated in the original SFT background (the NN string).

As we said above, based on the operator-state correspondence, we can assume that
in
the $c=1$ CFT a primary $\sigma_\ast$ exists
of weight $\frac 1{16}$ (but nonlocal in the conventional sense, \cite{Cardy}),
such that $\sigma_\ast(0)|0\rangle= |\sigma_\ast\rangle$, where $|0\rangle$ is
the vacuum for the (integral) $\alpha_n$ oscillators. However this is
not yet enough to justify the EM rules. {\it There is no guarantee that the
field $\sigma_\ast$
admits a free field representation in terms of the free field X or, what is the
same, in terms of the free integral oscillators $\alpha_n$. If it is true, it
must be proven. But if this is
not the case we have to give a sense to the operations of applying the SFT
operator $Q$ or string field $K$ (which we recall are expressed in terms of
free (integral-mode) matter and ghost oscillators) to
$\sigma_\ast,\bar\sigma_\ast$. }

To summarize we therefore face two alternatives.
In the best option we can express the fields $\sigma_\ast, \bar \sigma_\ast$
in terms of the $\alpha_n$ oscillators. In the worst, if this is not possible,
we must find a way to deal with different Hilbert spaces. The viability of such
options is far from obvious and in any case it has not yet been proved.

\section{Two alternatives}

The aim of this section is to find the correct
$K,B,c,\sigma_\ast, \bar\sigma_\ast$ algebra to justify the EM ansatz.
As said above, we have two kinds of Hilbert spaces,
one built out of the integral-mode matter oscillators $\alpha_n$ in the 25-th
direction, the
other with half-integral-mode oscillators
$J_{r}$, beside all the other matter and ghost oscillators.
Let us call them (or, better, their extensions) ${\cal H}$ and  ${\cal H}^\ast$,
respectively. The vacuum of the first is the usual string vacuum, made of the
tensor product of the various types of oscillators, in particular of the
$|0\rangle$ vacuum for the
integral oscillators $\alpha_n$, with the well-known star product of SFT. The
vacuum of the second is obtained by replacing the first vacuum
$|0\rangle$ with the state $|\sigma_\ast\rangle$ (or  $|\bar\sigma_\ast\rangle$)
on which the half-integer
oscillators act. In the first (extended) Hilbert space we have the usual $K,B,c$
algebra and the
BRST operator $Q=Q_{\rm ii}$ acting on it. In the second
$K$ is replaced by $K^\ast\equiv K_{\rm ND}\equiv K_{\rm DN}$ and $Q$ by
$Q^\ast\equiv Q_{\rm ND}\equiv Q_{\rm DN}$ and the star product is also modified accordingly.

Let us try to express this in formulas.
In the half-integral-mode space ${\cal H}^\ast$ with one ND or DN direction, the
BRST charge is given by
\begin{align}
Q^\ast=\sum_{n}c_nL_{-n}^{*}+\sum_{m,n}\frac{m-n}2:c_mc_nb_{-m-n}:-c_0\label{Qast}
\end{align}
where
\begin{align}
L_n^{*}={\cal L}_n+L_{n}^{(d-1)}\0
\end{align}
where $L_{n}^{(d-1)}$ is the contribution from the NN $X^{i}$'s,$
(i=0,\cdots 24)$ and ${\cal L}_n$ is as defined above. Instead, for the fully
integral-mode space ${\cal H}$, the BRST charge is
\begin{align}
Q=\sum_{n}c_nL_{-n}+\sum_{m,n}\frac{m-n}2:c_mc_nb_{-m-n}:-c_0\label{Qnotast}
\end{align}
We note that, since the Virasoro generators ${\cal L}_n$ obey the the same
algebra as $L_n$, the proof of the nilpotency of $Q^\ast$ is the same as that
of $Q$.

The vacuum state for the half-integral-mode space can be written as
\begin{align}
|0\rangle_{d}=|\sigma_\ast\rangle\otimes|0\rangle_{d-1}\otimes|0\rangle_{gh},
\end{align}
where $|\sigma_\ast\rangle$ is as defined above and
\begin{align}
&L_n^{(d-1)}|0\rangle_{d-1}=0,\quad {\rm for}~n\ge -1\0\\
&c_{n}|0\rangle_{gh}=0,\quad {\rm for}~n\ge 2,\qquad b_{n}|0\rangle_{gh}=0,\quad
{\rm for}~n\ge -1.
\end{align}
We also note that
\begin{align}
&{\cal L}_n|\sigma_\ast\rangle=0,\quad {\rm for}~n\ge 1.
\end{align}

Therefore, the action on the vacuum of the BRST  operator
$Q^\ast$ is well defined and is given by
\begin{align}
Q^\ast|0\rangle_{d}&=\sum_{n}c_{n}{\cal
L}_{-n}|\sigma_\ast\rangle\otimes|0\rangle_{d-1}\otimes|0\rangle_{gh}\0\\
&={\cal L}_{-1}|\sigma_\ast\rangle\otimes|0\rangle_{d-1}\otimes
c_1|0\rangle_{gh}+{\cal L}_{0}|\sigma_\ast\rangle\otimes|0\rangle_{d-1}\otimes
c_0|0\rangle_{gh}\0\\
&={\cal L}_{-1}|\sigma_\ast\rangle\otimes|0\rangle_{d-1}\otimes
c_1|0\rangle_{gh}+\frac1{16}|\sigma_\ast\rangle\otimes|0\rangle_{d-1}\otimes
c_0|0\rangle_{gh}\0
\end{align}
Translated into the BRST variation of the primary operator $\sigma_\ast$(s),
this
writes
\begin{align}
Q^\ast\sigma_\ast(s)=c(s)\partial\sigma_\ast(s)+\frac1{16}\partial
c(s)\sigma_\ast(s).
\end{align}

Now let us introduce $\sigma(s)=\sigma_\ast(s) e^{\frac i4X^0(s)},\quad
\bar\sigma(s)=\bar\sigma_\ast(s) e^{-\frac i4X^0(s)}$, which are the conformal
dimension zero bcc operators used in the EM ansatz, and use the shortcut
notation
\begin{align}
 \sigma(s)=\sigma_\ast(s) \omega(s),\quad \bar\sigma(s)=\bar\sigma_\ast(s)
\bar\omega(s).
\end{align}
The corresponding state $|\sigma\rangle$ is defined as follows
\begin{align}
|\sigma\rangle=|\sigma_\ast\rangle\otimes|\omega\rangle\otimes|0\rangle_{gh},
\end{align}
where $|\omega\rangle=\omega(0)|0\rangle_{d-1}$. The variation of this state
with respect $Q^\ast$ is
\begin{align}
Q^\ast|\sigma\rangle&=\sum_{n}{\cal
L}_{-n}|\sigma_\ast\rangle\otimes|\omega\rangle\otimes
c_{n}|0\rangle_{gh}+\sum_{n}|\sigma_\ast\rangle\otimes
L_{-n}^{(0)}|\omega\rangle\otimes c_{n}|0\rangle_{gh}\0\\
&={\cal L}_{-1}|\sigma_\ast\rangle\otimes|\omega\rangle\otimes
c_1|0\rangle_{gh}+\frac1{16}|\sigma_\ast\rangle\otimes|\omega\rangle\otimes
c_0|0\rangle_{gh}\0\\
&+|\sigma_\ast\rangle\otimes L^{(0)}_{-1}|\omega\rangle\otimes
c_1|0\rangle_{gh}-\frac1{16}|\sigma_\ast\rangle\otimes|\omega\rangle\otimes
c_0|0\rangle_{gh}\0\\
&={\cal L}_{-1}|\sigma_\ast\rangle\otimes|\omega\rangle\otimes
c_1|0\rangle_{gh}+|\sigma_\ast\rangle\otimes L^{(0)}_{-1}|\omega\rangle\otimes
c_1|0\rangle_{gh}\0
\end{align}
Translated into the BRST variation of the primary operator $\sigma(s)$, this is
written as
\be
Q^\ast\sigma(s)=c(s)\partial\sigma_\ast(s)\omega(s)+c(s)\sigma_\ast(s)\partial
\omega(s)=c(s)\partial\sigma(s).\0
\ee
A similar relation also holds for $\bar\sigma$.

So the action of $Q$ and $Q^\ast$ on the relevant vacua are well defined.
But it is easy to see that distingushing between ${\cal H}$ and ${\cal H}^\ast$, $K$ and $K^\ast$ is not
enough. Expressions like (\ref{sol}) are products of string states belonging to
different Hilbert spaces
\be
\Psi_{\rm NN} \Psi_{\rm ND} \Psi_{\rm DD} \Psi_{\rm DN} \Psi_{\rm NN}\label{NNXNN}
\ee
The first, third and fifth state belong to spaces of type ${\cal H}$, the
remaining ones to spaces of type ${\cal H}_\ast$. But
to be accurate we must introduce four Hilbert spaces ${\cal H}_{\rm NN}$,
${\cal H}_{\rm DD}$, ${\cal H}_{\rm ND}$ and ${\cal H}_{\rm DN}$. We tensor them with the
remaining matter
and ghost sectors and call them with the same name. The corresponding $Q$ and
$K$ will be labeled
in the same way, while $c$ and $B$ are universal. In particular $Q_{\rm NN}$ and
$Q_{\rm DD}$ are
(on zero momentum states) of type $Q$, and  $Q_{\rm ND}$ and $Q_{\rm DN}$ are of type
$Q^\ast$.

This poses several problems we have not considered so far.
\begin{itemize}
\item  what is the star product between string states belonging to different
Hilbert spaces?
\item the SFT BRST operator $Q=Q_{\rm NN}$ acts as a derivation on any expression
like (\ref{NNXNN}): how comes that it may change by taking the form appropriate
to different Hilbert spaces?
\end{itemize}
The answer to the first question does not seem to be unsurmountable in view of
the definition of star
product in terms of three-strings vertex. The second problem is more
complicated. For instance, how is it
possible that $Q$ becomes of type $Q^\ast$ whenever it comes across a factor
belonging to ${\cal H}^\ast$
and returns back to $Q$ if the subsequent factor belongs to ${\cal H}$? To
justify it we have envisaged two
possible alternatives.

\vskip 1cm
{\bf First alternative}. A way to realize this is by
introducing intertwining operators between the different Hilbert
spaces. Let us call them $X_{\rm NN,ND}, X_{\rm ND,DD},X_{\rm DD,DN}$ and $X_{\rm DN,NN}$ (we think
the names are self-explanatory). Next we rewrite (\ref{NNXNN}) as follows
\be
\Psi_{\rm NN}X_{\rm NN,ND} \Psi_{\rm ND}  X_{\rm ND,DD}\Psi_{\rm DD}X_{\rm DD,DN} \Psi_{\rm DN}X_{\rm DN,NN}
\Psi_{\rm NN}\label{NNXXNN}
\ee
and the relation among the different BRST operators is assumed to be
\be
&&Q_{\rm NN} X_{\rm NN,ND}=X_{\rm NN,ND}Q_{\rm ND}, \quad\quad Q_{\rm ND}
X_{\rm ND,DD}=X_{\rm ND,DD}Q_{\rm DD},\label{QXXQ}\\
&&Q_{\rm DD} X_{\rm DD,DN}=X_{\rm DD,DN}Q_{\rm DN}, \quad\quad Q_{\rm DN} X_{\rm DN,NN}=X_{\rm DN,NN}Q_{\rm NN}\0
\ee
Since any state $K$ is generated by the corresponding $Q$ acting on $B$ we have
also to assume
\be
&&K_{\rm NN} X_{\rm NN,ND}=X_{\rm NN,ND}K_{\rm ND}, \quad\quad K_{\rm ND} X_{\rm ND,DD}=X_{\rm ND,DD}K_{\rm DD},\label{KXXK}\0\\
&&K_{\rm DD} X_{\rm DD,DN}=X_{\rm DD,DN}K_{\rm DN}, \quad\quad K_{\rm DN} X_{\rm DN,NN}=X_{\rm DN,NN}K_{\rm NN}
\ee
If these relations are valid and we have the following modified version of the OPE \eqref{OPE2}
 \begin{align}\label{OPE3}
 &\big(X_{\rm DD,DN}\,\bar\sigma \,X_{\rm DN,NN}\big)\big(X_{\rm NN,ND}\,\sigma\, X_{\rm ND,DD}\big)=1,
 \end{align}
some of the issues raised in section 3 are resolved and we can prove the equation of motion and
also compute the energy, provided that the $Xs$ are pure matter operators (see appendix A).

Of course all the above works if the intertwining $X$ operators exist. Do they?
The possibility of
an intertwining operator is envisaged in \cite{Dixon}, accompanied by the
sentence ``The operator so constructed is rather unwieldy". But let us see what
these
authors refer to. The construction goes back to the 70's and is due to
E.Corrigan
and D. Fairlie, \cite{Corrigan}, but
see also \cite{Olive} for the environment (the search for off-shell
dual amplitudes)
where such an idea was born. Bcc operators are not mentioned, but they are
precisely what ref.\cite{Corrigan} deals with. More precisely that paper is
concerned
with the relation between what we call $|0\rangle\equiv |0\rangle_{25}$ and
$|\sigma_\ast\rangle$, and
defines
an intertwining operator for the vertex operators in the two pictures (integer
and half-integer
mode oscillators), say $V(k,z)$ constructed with $\alpha_n$ and $\tilde V(k,z)$
constructed
with the half-mode ones ($k$ is the momentum). The basic formula is
\be
\langle 0| \exp{F(w)}|\sigma_\ast\rangle\, V(k,z) = \lambda e^{-ik\cdot x_0}\,
\tilde V(k,z-w)  \,
\langle 0| \exp{F(w}|\sigma_\ast\rangle\label{inter}
\ee
where the quadratic form $F$ is given by
\be
F(z) = \frac 12 \frac 1{(2\pi i)^2} \oint_C \frac {dxdy}{xy} P(x) A(x,y,z)P(y)
+
\frac 1{(2\pi i)^2} \oint_{C'} \frac {dxdy}{xy} P(x) B(x,1/y,z)S(y)\0
\ee
where $P(z)= -z \frac d{dz} X_{\rm NN}(z)$ and $S(z)=-z\, J_ {\rm ND}(z)$, while
\be
A(x,y,z) = \sum_{n,m=0}^\infty x^n A_{n,m}(z) y^m = 2\log(\sqrt{x-z}+
\sqrt{y-z}), \quad\quad \lambda=2^{k^2}\label{A}
\ee
and
\be
B(x,1/y,z)= \sum_{n=0}^\infty \sum_{r=\frac 12}^\infty x^n B_{n,r}(z) y^{-r}=
\log\left( \frac{\sqrt y -\sqrt{x-z}}{\sqrt y +\sqrt{x-z}}\right)\label{B}
\ee
The integration contours must be such that for $C$ $|x|$ and $|y|<|z|$,  while
for $C'$ $|x|<|z|$ but $|y|>|x-z|$.
The intertwining operator $\langle 0| \exp{F(w)}|\sigma_\ast\rangle $ works very
well for vertex operators,
but it is easy to see that, unfortunately, it cannot work for expressions
like  $K_1^L$ and
$Q$. This example gives an explicit idea of what we mean by an intertwining
operator in the present context. We do not know whether this construction can be
improved, so as
to intertwine also $K_{\rm NN}$ and $Q_{\rm NN}$ with $K_{\rm ND}$ and $Q_{\rm ND}$, and to
intertwine
the star products in the two sector.
We do not exclude it, but our attempts in this direction have failed.

The idea of considering two different Fock spaces ${\cal H}$ and
${\cal H}^\ast$ with an intertwining operator between the them, if viable,
would
solve the problem. However such a construction
would make explicit use of DD, ND and DN oscillators, beside the NN ones.
The former appear in strings attached to a D-24 brane. In fact they define
the D24 brane (the dynamics of a brane is defined by the strings attached to
it).
This means that the information we want the solution to contain, i.e.
the description of the D24-brane, is already contained in the initial data. This
implies that {\it the solution is background dependent}.
There is nothing wrong, of course, in trying to describe D-branes
in a background dependent way. Background dependence is standard
in ordinary string theory approaches, for instance in \cite{GNS}.
But here we are in SFT and, as explained above, the ambition of this theory
is background independence.

This is a good point to recall that the non-analytic or approximate lump
solutions one finds
in the literature do not make use of bcc operators or half-integer modes. This
is the case for numerical solutions based on level truncations, \cite{numlump},
but also for the lump solutions in vacuum SFT (see, for instance, \cite{RSZ})
and for those in boundary SFT (see, for instance \cite{Kutasov}).
In analogy we expect that an analytic lump solution should not contain any
built-in information about D-branes, but rather a D-brane description
should emerge from the physical content of the solution. A comparison
with the BMT proposal below may help to understand the difference.

\vskip 1cm
{\bf Second alternative}. A radically different and more appealing alternative
could be inspired by the example of marginal deformation to the TV solutions.
To this end let us recall the KOS solutions,
\cite{KOS}.

\subsection{The KOS bcc operator}

In the case of the KOS solution for marginal deformations, the bcc operators
$\sigma$ and $\bar \sigma$ are
such that
\begin{align}\label{V-op}
\sigma(0)\bar\sigma(\alpha)={\rm exp}\Big[\int_0^\alpha dt V(t)\Big]
\end{align}
where $V$ is a matter primary operator of conformal dimension 1 and it belongs
to the original Hilbert space. Therefore, the bcc operators
belong to the same Hilbert space.  $V$ has the following properties
\begin{align}
[B,V]=[c,V]=0,\qquad QV=[K,cV].
\end{align}
The wedge states with the modified boundary condition are given by
\begin{align}\label{def-wedge}
e^{\alpha(K+ V)}=\sigma e^{\alpha K}\bar\sigma.
\end{align}
The BRST variation of the deformed wedge state is
\begin{align}
Qe^{\alpha(K+V)}&=\int_0^{\alpha}dte^{t(K+ V)}Q(K+ V)
e^{(\alpha-t)(K+V)}\0\\
&=e^{\alpha(K+ V)}( cV)- ( cV)e^{\alpha(K+ V)}\0
\end{align}
On the other hand using \eqref{def-wedge} we can write
\begin{align}
Qe^{\alpha(K+\ V)}=(Q\sigma) e^{\alpha K}\bar\sigma+\sigma e^{\alpha
K}(Q\bar\sigma).\0
\end{align}
Comparing the last two equations the authors of \cite{KOS} conclude that
\begin{align}
e^{\alpha(K+ V)}( cV)=\sigma e^{\alpha K}(Q\bar\sigma),\quad
- ( cV)e^{\alpha(K+ V)}=(Q\sigma) e^{\alpha K}\bar\sigma\0
\end{align}
Setting $\alpha=0$ and multiplying the first relation by $\bar \sigma$ from the
left and the second one
by $\sigma$ from the right we obtain
\begin{align}
Q\bar\sigma= c\bar\sigma V,\quad\label{Qsigma}
\sigma=- cV\bar\sigma,
\end{align}
We see that the original BRST operator operates effectively on the bcc
operators of the KOS solutions, because they belong to the original Hilbert
space.

The KOS solutions are based essentially on eq.(\ref{V-op}). Returning to the EM
ansatz, the
first difference we notice is that in the latter the field analogous to $V$ does
not exist (in the relevant case it
is singular, \cite{EM}). Thus we have to proceed in another way. In
\cite{Zamolo} the half-integral-mode
sector is called `Ramond'. This is an unconventional terminology, but it
is reminiscent of the Ramond
sector in superstring theories and suggests a parallel with that situation. We
recall
that in open superstring theory we have two vacua, the NS one, analogous to
the usual vacuum in bosonic string theory, and the Ramond vacuum (which is made
of spinor states forming a representation of the gamma matrix algebra in 10D).
It is however possible to define a primary operator that, applied to the usual
vacuum, creates the Ramond vacuum under the state-operator correspondence. This
primary
is constructed using (matter and ghost) fields in the theory (with fermion
and superghost fields in bosonized form).

Applying this analogy to the EM proposal we may ask whether the
$|\sigma_\ast\rangle$ (and $|\bar\sigma_\ast\rangle$) vacuum can be created in an analogous way starting from
the ordinary vacuum and
applying a primary operator formed with the fields in the theory. In such a case
all $Q$'s and $K$'s would
collapse to the same operator and the previous difficulties would disappear,
because
there would be no need to distinguish between string states belonging to the four different Hilbert spaces.
This scheme is certainly
more appealing but not easy to implement. It cannot be done straight away,
because we have at our disposal only the bosonic NN field $X$ (which has no
charge
at infinity, and thus does not offer any chance to use the Coulomb gas
method\footnote{Needless to say $e^{\frac 14 X}$, where $X$ is as in (\ref{XN}),
is not what we need.}).
At page 5 of \cite{Hashimoto} we find the peremptory sentence:
``In open strings there is no such
thing as a twisted state", where `twisted state' refers to
$|\sigma_\ast\rangle$. Perhaps this is too strong a statement, but, certainly,
there seems to be no straightforward way to implement this scheme.

{\it In conclusion we have not been able to implement the EM prescriptions in a
concrete 2D field theory formalism,} consistent with 2D CFT on which SFT is
defined.
But, since ours is not a mathematical theorem, we cannot completely exclude
that it is possible.

The parallel with the KOS solution suggests one more consideration. When one
tries to translate the KOS solution into the EM scheme one comes across a
singularity (in the
analogue of $V$). This may be
an indication that lump solutions are inevitably, in some sense, singular. A
similar peculiarity
is met in numerous classical field theory solutions and, in itself, is not
really important.
What is important in these cases is that the physical quantities related to the
solutions can be computed. This is what happens also for the BMT solution.

\section{The BMT proposal}
\label{sec:BMT}

In \cite{BMT}{} a general method has been proposed to obtain new
exact analytic solutions in open string field theory, and in particular
solutions that describe inhomogeneous tachyon condensation. The method consists
in translating an exact renormalization group (RG) flow generated in a
two--dimensional world--sheet theory by a relevant operator, into the language
of OSFT. The so-constructed solution is a deformation of the ES solution.
It has been shown in \cite{BMT} that, if the operator has
suitable properties, the solution will describe tachyon condensation only in
specific
space directions, thus representing the condensation of a lower dimensional
brane. In the following, after describing the general method, we will focus on a
particular
solution, generated by an exact RG flow first analyzed by Witten
\cite{Witten}{}.
 On the basis of the analysis carried out in the framework of 2D CFT in
\cite{Kutasov}{}, we expect it to describe a D24-brane, with the correct ratio
of tension with respect to the starting D25 brane.

Let us see first the general recipe to construct such kind of lump solutions. To
start with we enlarge the
$K,B,c$ algebra by adding a (relevant) matter string field $\phi$, where
\begin{eqnarray}
\phi=\phi\left(\frac12\right)\ket I,\quad\quad c=c\left(\frac12\right)\ket I \label{phi},\0
\end{eqnarray}
with the properties
\begin{eqnarray}
\,[c,\phi]=0,\quad\quad \,[B,\phi]=0,\quad\quad \,[K,\phi]= \del\phi,\quad\quad
Q\phi=c\del \phi+\del c\delta\phi. \label{actionQ}
\end{eqnarray}
It can be easily proven that
\begin{eqnarray}
\psi_{\phi}= c\phi-\frac1{K+\phi}(\phi-\delta\phi) Bc\del c\label{psiphi}
\end{eqnarray}
does indeed satisfy (in the sense specified below) the OSFT equation of motion
\begin{eqnarray}
Q  \psi_{\phi}+\psi_{\phi}\psi_{\phi}=0\label{eom}
\end{eqnarray}
It is clear that (\ref{psiphi}) is a deformation of the Erler--Schnabl solution,
which can be recovered for $\phi=1$.

After some algebraic manipulations one can show
that
\begin{eqnarray}{\cal Q}_{\psi_\phi} \frac B{K+\phi}=Q\frac
B{K+\phi}+\left\{\psi_\phi,\frac B{K+\phi}\right\}=1.\0
\end{eqnarray}
So, unless
the string field $\frac B{K+\phi}$ is singular, it defines a homotopy operator
and the solution has trivial cohomology, which is the defining property of the
tachyon vacuum
\cite{Ellwood}{}. On the other hand, in order for the solution to be
well defined, the quantity $\frac1{K+\phi}(\phi-\delta\phi)$ should
be well defined. Moreover, in order to be able to show that (\ref{psiphi})
satisfies the equation of motion, one needs $K+\phi$ to be invertible.

In full generality we thus have a new nontrivial solution if, roughly speaking,
\begin{enumerate}
\item $\frac1{K+\phi}$ is in some sense singular, but
\item the RHS of (\ref{psiphi}) is regular and
\item $\frac1{K+\phi}(K+\phi)=1$
\end{enumerate}
These conditions seem hard to satisfy and even contradictory. It is indeed so without adequate
specifications. This problem was discussed in \cite{BGT1,BG}, where it was shown
that the right framework for a correct interpretation is distribution theory, which guarantees not only
regularity of the solution but also its `non-triviality', in the sense that if
these conditions are satisfied, it cannot fall in the same class as the ES
tachyon vacuum solution. These questions will be discussed further on.

For concreteness the worldsheet RG flow, referred to above, is represented by
a parameter $u$, where $u=0$ corresponds to the UV and
$u=\infty$ to the IR (in 2D), and $\phi$ is labelled as $\phi_u$, with $\phi_{u=0}=0$.
Then we require
for $\phi_u$ the following properties under the coordinate rescaling
$f_t(z)=\frac zt$
\begin{eqnarray}
f_t\circ\phi_u(z)=\frac1t\,\phi_{tu}\left(\frac zt\right).\label{cnd1}
\end{eqnarray}

In \cite{BMT} a specific relevant operator $\phi_u$ and the
corresponding SFT solution was considered. This operator generates an exact RG
flow studied by Witten in
\cite{Witten}{}, see also \cite{Kutasov}{}, and is based on the operator
(defined in the cylinder $C_T$ of width $T$ in the arctan frame)
\begin{eqnarray}
\phi_u(s) = u (X^2(s)+2\,\ln u +2 A)\label{TuCT1}
\end{eqnarray}
where $X=X^{25}$ is the NN open string field and $A$ is a constant.
On the unit disk $D$ we have
\begin{eqnarray}
\phi_u(\theta) = u (X^2(\theta)+ 2\,\ln \frac {Tu}{2\pi} +2 A)\label{TuDb1}
\end{eqnarray}

If we set
\begin{eqnarray}
g_{A}(u) =\langle e^{- \frac 1{2\pi } \int_{0}^{4{\pi}}
d\theta \, u\Bigl{(} X^2(\theta) + 2 \ln \frac u{2\pi}+2 A\Bigr{)}}\rangle_{D}
\0
\end{eqnarray}
we get
\begin{eqnarray}
g_{A}(u) = Z(2u)e^{-2u (\ln \frac u{2\pi}+A)}\label{gAu2}
\end{eqnarray}
where $Z(u)$ is the partition function of the system on the unit disk computed
by \cite{Witten}{}.
Requiring finiteness for $u\to\infty$ one gets $A= \gamma -1+\ln 4\pi$, which
implies
\begin{eqnarray}
g_{A}(u)\equiv g(u)= \frac 1{2\sqrt{\pi}} \sqrt{2u} \Gamma(2u) e^{2u(1-\ln
(2u))}, \quad\quad
\lim_{u\to\infty} g(u) = 1\label{inflimitgA1u}
\end{eqnarray}
Moreover, as it turns out, $\phi_u -\delta\phi_u = u \del_u \phi_u(s)$

The $\phi_u$ just introduced satisfies all the requested properties.
According to \cite{Kutasov}{}, the corresponding RG flow in BCFT reproduces the
correct ratio of tension between D25 and D24 branes.
Consequently $\psi_u\equiv \psi_{\phi_u}$ is expected to represent a D24 brane
solution.

In SFT the most important gauge invariant quantity is of course the energy.
Therefore in
order to make sure that $\psi_u\equiv \psi_{\phi_u}$ is the expected solution we
must prove
that its energy equals a D24 brane energy.

The energy expression for the lump solution was determined in \cite{BMT} by
evaluating a three--point
function on the cylinder $C_T$. It equals $-\frac 16$ times the following
expression
\begin{eqnarray}
&&\langle \psi_u \psi_u\psi_u\rangle=-\int_{0}^\infty
dt_1dt_2dt_3{\cal
E}_0(t_1,t_2,t_3)u^3g(uT)\Bigg\{\Big(-\frac{\partial_{2uT}g(uT)}{g(uT)}
\Big)^3\0\\
&&~~~~~~+\frac
12\Big(-\frac{\partial_{2uT}g(uT)}{g(uT)}\Big)\Big(G_{2uT}^2(\frac{2\pi
t_1}T)+G_{2uT}^2(\frac{2\pi (t_1+t_2)}T)+G_{2uT}^2(\frac{2\pi
t_2}T)\Big)\0\\
&&~~~~~~+G_{2uT}(\frac{2\pi t_1}T)G_{2uT}(\frac{2\pi
(t_1+t_2)}T)G_{2uT}(\frac{2\pi t_2}T)\Bigg\}\label{LLL}
\end{eqnarray}
Here $T=t_1+t_2+t_3$ and $g(u)$ is as above, while $G_u(\theta)$ represents the
boundary-to-boundary correlator first determined by Witten\cite{Witten}{}:
\begin{eqnarray}
G_u(\theta)= \frac {1}{u} +2 \sum_{k=1}^\infty \frac {\cos (k\theta)}{k+u} \0
\end{eqnarray}
Finally ${\cal E}_0(t_1,t_2,t_3)$ represents the ghost three--point function in
$C_T$.
\begin{eqnarray}
{\cal E}_0(t_1,t_2,t_3)=\left\langle Bc\del c(t_1+t_2)\del c(t_1) \del c(0)
\right\rangle_{C_T}= -\frac 4{\pi} \sin \frac {\pi t_1}T \sin \frac
{\pi(t_1+t_2)}T \sin \frac{\pi t_2}T\0
\end{eqnarray}
A remarkable property of (\ref{LLL}) is that it does not depend on $u$. In fact
$u$ can be absorbed in a redefinition of variables $t_i \to ut_i$, $i=1,2,3$,
and disappears from the expression.

The integral in (\ref{LLL}) is well defined in the IR ($s$ very large, setting
$s=2uT$) but has
a UV ($s\approx 0$) singularity, which must be subtracted away\footnote{Of
course this singularity
is present also in the ES solution, it is represented by the infinite volume.
The difference
here is that the infinite volume appears in the form of a zero mode which
generates
a singularity $\sim \frac 1{\sqrt u}$ in $g(u)$.}.
Once this is done, the expression (\ref{LLL}) can be numerically computed, the
result being $\approx 0.069$. This is not the expected result, but this is not
surprising,
for the result depends on the UV subtraction. Therefore one cannot
assign to it any physical significance. To get a meaningful result
we must return to the very meaning of Sen's third conjecture, which says that
{\it the lump solution is a solution of the theory on the tachyon condensation
vacuum}. Therefore we must measure the energy of our solution with respect to
the tachyon condensation vacuum. Simultaneously the resulting energy must be a
subtraction-independent quantity because only to such a
quantity can a physical meaning be assigned.
Both requirements have been satisfied in \cite{BGT1} in the following way.

First a new solution to the EOM, depending on a parameter $\ve$, has been
introduced
\begin{eqnarray}
\psi_u^{\ve}= c(\phi_u+\ve) - \frac 1{K+\phi_u+\ve} (\phi_u+\ve -\delta \phi_u)
Bc\partial c.
\label{psiphieps}
\end{eqnarray}
in the limit $\ve\to 0$. This limit will be mostly understood from now on.
The energy of (\ref{psiphieps}) (after the same UV subtraction as in the
previous case\footnote{The parameter $\varepsilon$ is originally a gauge parameter,
but due to the UV subtraction such gauge nature is broken and the energy functional
depends softly on the value of $varepsilon$, \cite{BGT3}.}) is (numerically) 0. Since (unlike the previous case) the presence
of the parameter $\ve$ prevents the IR transition to a new critical point, it
seems
sensible to assume that $\lim_{\ve\to 0} \psi_u^{\ve}$ represents the tachyon
condensation vacuum solution. In other words we assume it is gauge equivalent to
the ES
solution (we will justify this further on).
Using it, {\it a solution to the EOM at the tachyon condensation vacuum} has
been obtained. The equation of motion at the tachyon vacuum is
\begin{eqnarray}\label{EOMTV}
{\cal Q}\Phi+\Phi\Phi=0,\quad {\rm where}~~{\cal
Q}\Phi=Q\Phi+\psi_u^\ve\Phi+\Phi\psi_u^\ve.
\end{eqnarray}
One can easily show that
\begin{eqnarray}\label{psiupsie}
\Phi_0=\psi_u-\psi_u^\ve
\end{eqnarray}
is a solution to (\ref{EOMTV}). The action at the tachyon vacuum is
$-\frac12\langle{\cal
Q}\Phi,\Phi\rangle-\frac13\langle\Phi,\Phi\Phi\rangle.$ Thus the energy of
$\Phi_0$ is
\begin{eqnarray}\label{TVlumpenergy}
E[\Phi_0]&=&-\frac16\langle\Phi_0,\Phi_0\Phi_0\rangle\0\\
&=&
-\frac16\big[\langle\psi_u,\psi_u\psi_u\rangle-\langle\psi_u^\ve,
\psi_u^\ve\psi_u^\ve\rangle
-3\langle\psi_u^\ve,\psi_u\psi_u\rangle+3\langle\psi_u,
\psi_u^\ve\psi_u^\ve\rangle\big].\label{EPhi0}
\end{eqnarray}
The UV subtractions necessary for each correlator at the RHS of this equation
are the same in all cases, therefore they
cancel out and the final result is subtraction-independent. A final bonus of
this procedure is that the final result can be derived purely analytically and
$E[\Phi_0]$ turns out to be precisely the D24-brane energy. With the conventions
of \cite{BGT1}, this is
\begin{eqnarray}
T_{D24}= \frac 1{2\pi^2}\label{T24}
\end{eqnarray}
In \cite{BGT2} the same result was extended to Dp-brane lump solutions for any
$p$. All these solutions are background independent.

Before we pass to a closer scrutiny of this solution and its properties, we
need a discussion of the relations
between various solutions of the SFT eom provided by gauge transformations.
What we would like to stress is that gauge
equivalent solutions may take very different (even singular) forms. To this end
we make a detour about identity based solutions (IB).

\section{A detour: TV solution as gauge transformation of IB solutions}

Let us introduce the following family of solutions, referred to as identity
based
(IB) \cite{Kishimoto:2009nd,Zeze:2010jv,Zeze:2010sr,Arroyo:2010sy}, depending on
a parameter $\alpha$:
\be
\psi_\alpha=c(\alpha-K),\label{IB}
\ee
which can also be written as a pure gauge solution as follows
\be
\psi_\alpha=U_\alpha^{-1}QU_\alpha,\label{IB2}
\ee
where
\be
U_\alpha=1-\frac1\alpha cB(\alpha-K),\qquad
U_\alpha^{-1}=1+cB\frac{\alpha-K}K.\0
\ee
We note that for the $\alpha=0$ case the gauge transformation is singular. The
homotopy field is ${\cal A}_\alpha=\frac 1\alpha B$.
The problem with identity based solutions is in the computation of the physical
observable. For example, the direct evaluation of the energy using the $K,B,c$
algebra gives zero.
\be
E\sim \langle \psi_\alpha^3\rangle  =-\langle cKcKcK\rangle
=-\langle cKc\partial cK\rangle=-\langle cK(\partial c )^2K\rangle=0.\0
\ee
However, if we use different regularization method we will get different results.
The most obvious regularization is the following
  \be
\langle \psi_\alpha^3\rangle & =-\lim_{t_i\to0}\langle
cK\Omega^{t_1}cK\Omega^{t_2}cK\Omega^{t_3}\rangle=
\lim_{t_i\to0}\partial_{t_1}\partial_{t_2}\partial_{t_3}\langle
c\Omega^{t_1}c\Omega^{t_2}c\Omega^{t_3}\rangle.
\ee
The answer in this case depends on the way we take the limits, therefore this
regularization is ambiguous. Another procedure is to connect this
solution to the class of ES TV solutions by a regular gauge transformation
\cite{Zeze:2010jv,Zeze:2010sr,Arroyo:2010sy}. This
way we generate from the
identity based solutions the following two parameters family of solutions
\be
\Psi_{\alpha,\lambda}=V_\lambda^{-1}(\psi_\alpha+Q)V_\lambda=\frac1\lambda
c(1+\lambda KBc)\left(\frac 1{1+\lambda K}+\lambda\alpha-1\right)
\ee
where
\be\label{ES-type}
V_\lambda^{-1}=1+\lambda cBK,\qquad V_\lambda=1-\lambda cBK\frac1{1+\lambda K}.
\ee
If we set $\lambda=\alpha=1$, we obtain the ES TV solution. Since all the known
solutions
can formally be written as a pure gauge, for any solution
$\Psi=U_\Psi^{-1}QU_\Psi$ we can write
 \be
\Psi\equiv U_\Psi^{-1}QU_\Psi\equiv V_\Psi^{-1}(\psi_\alpha+Q)V_\Psi
=V_\Psi^{-1}(U_\alpha^{-1}QU_\alpha+Q)V_\Psi,
\ee
where
\be
V_\Psi=U_\alpha^{-1}U_\Psi.
\ee
For the TV solutions this gauge transformation is regular, which means they are
gauge equivalent to the IB solutions.

The energy for these class of solutions in \eqref{ES-type} is independent of
both $\alpha$ and $\lambda$ and it is equal to the energy of the ES TV solution.
The same is true for the closed string overlap. Since the original Schnabl's TV
solution is also gauge equivalent to the ES solution, we see that all the known
TV solutions are regular gauge transforms of the IB solutions.

What is interesting about the generalized TV solution
$\Psi_{\alpha,\lambda}$ is related to its homotopy field. The homotopy field is
given by
\be
{\cal A}_{\alpha,\lambda}=V_\lambda^{-1}{\cal A}_\alpha V_\lambda=\frac 1\alpha
\frac B{1+\lambda K},
\ee
which is a well defined field for $\alpha\ne0$ and $\lambda\ge0$. However, the
solution itself and the corresponding physical observables are well defined for
$\alpha=0$ as well. We can not tell if the cohomology of the corresponding
BRST operator is vanishing for $\alpha=0$. We know that the existence of
the homotopy field is a sufficient condition for the vanishing of the comoholgy,
but is it a necessary condition? We will return to this point later on.

\subsection{BMT lump solution as gauge transformation of IB solutions}

Let us apply the previous formalism to the BMT lump solution. In particular
we write the gauge transformation which relates the BMT lump solution to
the identity based solution. Since the parameter $\lambda$ has no significance
we set it to one.
\be
 \Psi_{\phi,\alpha}\equiv U_\phi^{-1}QU_\phi\equiv
V_\phi^{-1}(\psi_\alpha+Q)V_\phi,
\ee
 where $V_\phi=U_\alpha^{-1}U_\phi$, is given by
 \be
 V_\phi^{-1}=1+cB(K+\phi-1),\qquad V_\phi=1-cB\left(1-\frac1{K+\phi}\right).
 \ee
 After some simplifications we obtain
 \be\label{BMT-lump}
 \Psi_{\phi,\alpha}=\alpha\phi
c+(\alpha-1)cKBc-cKcB(\phi-\delta\phi)\frac1{K+\phi}.
 \ee
We note that for $\alpha=1$ this gives the $\psi_\phi$ BMT lump ansatz. The new
thing here is that this time the gauge transformation is not regular (see below for an
additional comment on regular and singular gauge trasformations).
Therefore, we can claim that the lump solution is a genuine new solution. The
calculation of physical observable will follow the standard procedure. For
example the first two pieces in \eqref{BMT-lump} do not contribute to the
energy, so that the energy is $\alpha$ independent and the same as that of the
BMT
lump solution.  Let us focus on the homotopy field. It is given by
\be
{\cal A}_{\phi,\alpha}=V_\phi^{-1}{\cal A}_\alpha V_\phi=\frac 1\alpha \frac
B{\phi+K},
\ee
which is singular even for $\alpha\ne 0$.

In \cite{BGT1} another regularized solution was used (see above). It was
obtained
via the replacement $\phi\to\phi+\varepsilon$. Replacing this in the above formulas
gives
\be
 V_\phi=1+cB(K+\phi+\ve-1),\qquad
V_\phi^{-1}=1-cB\left(1-\frac1{K+\phi+\ve}\right),\label{regsol}
 \ee
and the solution becomes
\be\label{BMT-lump-eps}
 \Psi_{\phi,\alpha,\ve}=\alpha(\phi+\ve)
c+(\alpha-1)cKBc-cKcB(\phi+\ve-\delta\phi)\frac1{K+\phi+\ve}.
\ee
The regularized solution is obtained with $\alpha=1$. In  \cite{BGT1} it was
identified with the TV solution (see also the previous section).
The gauge transformation this time is regular because of the $\varepsilon$
parameter, and the solution will
reproduce the TV observables. In this case the homotopy field is
 \be
 {\cal A}_{\phi,\alpha}=\frac 1\alpha \frac B{K+\phi+\ve},
 \ee
 which is regular as expected (see further comments below on this issue). The
regularized solution is also gauge equivalent
to the ES solution. Setting $\alpha=1$ we can write
 \be
 \Psi_{\phi,\ve}=X^{-1}(\Psi_{ES}+Q)X\label{gaugetransf1}
 \ee
 where
 \be
 X^{-1}=1-cB(1-\phi-\ve)\frac1{K+1},\qquad
X=1+cB(1-\phi-\ve)\frac1{K+\phi+\ve}\label{gaugetransf2}
 \ee
so it has precisely the same energy as the ES solution.

\section{Discussion of BMT solution}

This section is devoted to discussing the critical aspects of the BMT proposal.
Let us start with
the conditions 1,2 and 3 of section \ref{sec:BMT}. We have already anticipated
there that
these, seemingly contradictory, requirements can be satisfied with an
appropriate
mathematical interpretation. String states such as $\frac 1{K+\phi}$ or,
similarly, $\frac 1K$ may be singular. This is so
because the operators that act on $\ket I$ in the definition of both are
expected to have a nontrivial kernel.
In such a case the Schwinger representations
\be
\frac 1K= \int_0^\infty dt e^{-Kt}, \quad\quad \frac 1{K+\phi}= \int_0^\infty dt
e^{-(K+\phi)t}\label{schwinger}
\ee
are bound to diverge, due the zero modes of the corresponding operators. To
obtain well-behaved
Schwinger representations we have to find a way to remove such singularities.

As was repeatedly noticed in \cite{BGT1,BGT2,BG}, this is analogous to the
procedure of removing singularities,
or defining distributions, in ordinary function theory. In \cite{BG} a first
attempt was made to do the same
for $\frac 1{K+\phi}$. A space ${\boldsymbol {\EF}}$ of {\it test string states}
or {\it regular states}
with a definite topology was introduced.
The dual of it, ${\boldsymbol {\EF}'}$, with the appropriate topology was
defined. The state   $\frac 1{K+\phi}$ belongs to the
latter, i.e. it is a {\it distribution}. As such the condition 2 and 3 of
section \ref{sec:BMT} are satisfied and the
SFT equation of motion is verified (on the other hand, with a grain of good
sense, a simple continuity argument
would be enough to prove the latter, see \cite{BGT1}). Moreover, as was shown
above, the procedure to compute the energy is well
defined.

As for point 1 of sec.\ref{sec:BMT} it is self-evident, but a comment is in
order.
The string state
$\frac 1{K+\phi}$ must be singular so that the would-be homotopy operator $\frac
B{K+\phi}$ (understood as the star-multiplication
operator by the string state $\frac B{K+\phi}$) be not well-defined,
otherwise the perturbative spectrum on the brane would be trivial. A true
homotopy operator is supposed to map
normalized states (annihilated by the BRST operator) into normalized states. The
state $\frac 1{K+\phi}$ is singular
and it has to be regularized. As was shown in \cite{BG} this can be done in a
weak sense (in physical language,
for correlators), while the previous requirement would require an operator
topology argument, which does not seem to exist.
For this reason we conclude that  $\frac B{K+\phi}$ does not exist as a homotopy
operator.

Let us illustrate this with an example. In the concrete BMT solution in sec. 7,
defined by the relevant perturbation (\ref{TuCT1}), the inhomogeneous tachyon
condensation
takes place in the limit $u\to\infty$. Now it is easy to see that
\be
{\cal Q} \big(c\phi_u\big) = c K \frac 1{K+\phi_u} (\phi_u-2u)Bc\del
c\label{cphi}
\ee
and, naively,
\be
\lim_{u\to\infty} \frac 1{K+\phi_u} (\phi_u-2u)=1\0
\ee
Therefore $c\frac {\phi_u}u$ is a closed state for ${\cal Q}$ in the limit
$u\to\infty$,
but it is infinite due to the presence of $\log u$ in $\phi_u$. On the other
hand
one can easily show in the same way that $\lim_{u\to \infty} {\cal Q}c=0$. So
the (non-singular) state
\be
\phi^{(as)}=\lim_{u\to\infty} c \left( \frac {\phi_u}u-2u\log
u\right)\label{phias}
\ee
is annihilated by ${\cal Q}$ (at least to this naive level of manipulation).
Therefore, provided it is nontrivial, it is a candidate
to represent a (zero momentum) state in the spectrum of the D24-brane. Now,
while it is
not hard to imagine a procedure to define a norm for this state, it is
impossible to
do the same for the state $\frac B{K+\phi}\phi^{(as)} $.

The previous conclusions have important consequences for the themes discussed in
this paper.
For instance, it is debated what the allowed gauge transformations are in SFT.
Based on the above we propose the following distinction: a gauge
transformation is allowed if it is regular in the
above sense, that is if it does not need to be regularized; otherwise it is not
allowed.
In other words, an
allowed gauge transformation cannot be a true distribution.
Once this is established, we can return to the end of the previous section:
the state $\frac 1{K+\phi+\ve}$ for $\ve \neq 0$ is not a distribution (it does
not need a
regularization), it is a regular state. So the states $X$ and $X^{-1}$ that
leads from the ES solution to
the $\Psi_{\phi,\ve}$ solution, (\ref{gaugetransf1},\ref{gaugetransf2}), are
regular.
Therefore the corresponding
gauge transformation is a genuine one, and the $\Psi_{\phi,\ve}$ solution is
equivalent to
the ES one. This implies in
particular that they have the same energy.

We can now reconsider one of the unsatisfactory aspects of the energy
calculation in \cite{BGT1}.
As pointed out above, the energy of (\ref{regsol}) was calculated only
numerically and turned out to
be 0. It was remarked in  \cite{BGT1} and repeated above that this value is only
conventional, because obtained via
a UV subtraction which introduces an arbitrariness in the absolute result.
Consequently it was stressed
in \cite{BGT1} that only quantities that are independent of such subtractions
can be assigned a
physical meaning, and it was
precisely in this way that the lump energy was calculated. From what we have
just argued (see also the end of sec. 7.1)
we are now able not only to confirm all this, but also that the energy of
$\Psi_{\phi,\ve}$ is
precisely the TV energy, thus confirming the intuition in  \cite{BGT1}.

\subsection{A comment about homotopy operators}

In the previous section we have presented a family of analytic solutions of the
OSFT eom
that are gauge equivalent to identity based solutions. They split into two
subsets:
$\alpha\neq 0$ and $\alpha=0$.
The first set is gauge equivalent to the ES solution and has a regular homotopy
operator. The second set, although modelled on the ES solution, although regular
and with the same energy as the ES solution, is in no obvious way gauge
equivalent to it. The latter case is puzzling because, in addition, it does not
admit a homotopy operator as a function of $K,B,c$. We recall again that the
existence of a homotopy operator, in correspondence with a tachyon condensation
vacuum solution $\psi_0$, implies that that vacuum has trivial cohomology, i.e.
no perturbative open string spectrum. For ${\cal Q}_{\psi_0}\phi=0$  implies
\be
\phi = ({\cal Q}_{\psi_0}{\cal A})\phi= {\cal Q}_{\psi_0} ({\cal A}\phi)-
{\cal A}({\cal Q}_{\psi_0}\phi)= {\cal Q}_{\psi_0} \chi,\quad\quad \chi={\cal
A}\phi\0
\ee
This condition is also necessary: if any state annihilated by ${\cal
Q}_{\psi_0}$ is trivial,
i.e. if ${\cal Q}_{\psi_0} \phi=0$ implies that $\phi = {\cal Q}_{\psi_0}\chi$
for some state $\chi$, it means that there exists a map whose domain is the
kernel of ${\cal Q}_{\psi_0}$ and whose image lies in the complement of the
kernel. This map is linear and is precisely the homotopy operator\footnote{We do
not say anything about the properties of the homotopy operator, for instance if
it is continuous. To this end one should introduce a topology in the space of
string states. This problem is usually ignored in the existing literature,
although it is very likely at the origin of many ambiguities that arise in the
search for analytic solutions in SFT.}.

This is puzzling and we face two possibilities: (1) either $\psi_{\alpha=0}$ is
truly
not equivalent to the ES solution though degenerate in energy with it, in which
case it may well be that
the perturbative spectrum supported by this solution is not empty; but the
physical interpretation would be obscure: what is this vacuum degenerated with
the TV supposed to be? or (2) $\psi_{\alpha=0}$ is truly equivalent in some non
evident way to the ES solution, but in this case a homotopy operator should
exist. Then we are again faced with two possibilities. The first
possibility is that the homotopy operator for $\psi_{\alpha=0}$ exists, but
cannot be
expressed as a function $K,B,c$. Although unlikely, because the solution
$\psi_{\alpha=0}$ is a simple function of $K,B,c$, we cannot completely exclude
this
exotic possibility. A second, more likely, alternative is that the homotopy
operator exists as a limit of a sequence of homotopy operators of analogous
solutions. This of course requires the existence of a topology in the space of
solutions, that is in the space of string states. Such a topology should not be
confused with the Hilbert space norm, because most of the string states that
enter this game have infinite hilbert space norm. An attempt in this direction has
been initiated in \cite{BG}.

\section{Conclusions}

In this paper we have pointed out the problems connected with the proposal of
\cite{EM} to construct analytic lump solutions in OSFT. We have remarked that
for it to become effective (in a background independent way) one must show that
it is possible to implement
it in the very same language (2D CFT) in which OSFT is formulated. We have made
some attempts in this direction without succeeding. However we  do not
completely exclude that
such an implementation be possible. We have also made some additional improving
remarks on the
BMT proposal for analytic lump solutions. This proposal too has a problem of a
different nature related
to its mathematical background: a formulation of a well-defined distribution
theory for string fields
that generalizes the approach of \cite{BG}. We hope anyhow we have at least
brought enough evidence, with this
and other examples in section 8.1, that it is time to face the problem of the
topology in the space of string fields.

\vskip 1cm

{\bf Acknowledgements} We would like to thank Carlo Maccaferri for explaining to us his
work prior to publication and for other comments, and Ilmar Gaharamanov for collaboration
in an early stage of the research. The work of D.D.T. was supported by the
grant MIUR 2010YJ2NYW\!\_\!\_001.

\appendix
\section{EoM and Energy with intertwining operators}

In this appendix we verify the equation of motion and energy of the EM solution, rewritten in terms of intertwining operators as in section 5. Eq.\eqref{Phi0}  writes
\begin{align}\label{Phi02}
 \Phi_0=-\frac1{\sqrt{1+K_{\rm NN}}}c(1+K_{\rm NN})X_1\sigma X_2\frac B{1+K_{\rm DD}}X_3\bar\sigma X_4 (1+K_{\rm NN})c\frac1{\sqrt{1+K_{\rm NN}}},
 \end{align}
where we used
$X_1=X_{\rm NN,ND},\quad X_2=X_{\rm ND,DD},\quad X_3=X_{\rm DD,DN}$ and $X_4=X_{\rm DN,NN}$ for simplicity. Using \eqref{QXXQ}, the BRST variation of $\Phi_0$ is written as
\begin{align}
 Q_{\rm NN}\Phi_0&=-\frac1{\sqrt{1+K_{\rm NN}}}(Q_{\rm NN}c)(1+K_{\rm NN})X_1\sigma X_2\frac B{1+K_{\rm DD}}X_3\bar\sigma X_4 (1+K_{\rm NN})c\frac1{\sqrt{1+K_{\rm NN}}}\0\\
 &+\frac1{\sqrt{1+K_{\rm NN}}}c(1+K_{\rm NN})X_1(Q_{\rm ND}\sigma)X_2 \frac B{1+K_{\rm DD}}X_3\bar\sigma X_4 (1+K_{\rm NN})c\frac1{\sqrt{1+K_{\rm NN}}}\0\\
 &+\frac1{\sqrt{1+K_{\rm NN}}}c(1+K_{\rm NN})X_1\sigma X_2 \frac {(Q_{\rm DD}B)}{1+K_{\rm DD}}X_3\bar\sigma X_4 (1+K_{\rm NN})c\frac1{\sqrt{1+K_{\rm NN}}}\0\\
 &-\frac1{\sqrt{1+K_{\rm NN}}}c(1+K_{\rm NN})X_1\sigma X_2\frac B{1+K_{\rm DD}}X_3(Q_{\rm DN}\bar\sigma)X_4 (1+K_{\rm NN})c\frac1{\sqrt{1+K_{\rm NN}}}\0\\
 &-\frac1{\sqrt{1+K_{\rm NN}}}c(1+K_{\rm NN})X_1\sigma X_3\frac B{1+K_{\rm DD}}X_3\bar\sigma X_4(1+K_{\rm NN})(Q_{\rm NN}c)\frac1{\sqrt{1+K_{\rm NN}}}.\0
 \end{align}
Unlike what we see in equation \eqref{QPhi0}, now the switching of the BRST charge from one Hilbert space to the other when it acts on different components of the solution is justified by \eqref{QXXQ}.  Using the $K,B,c,\sigma,\bar\sigma$ algebra of section 3 we can write
\begin{align}
 Q_{\rm NN}\Phi_0&=-\frac1{\sqrt{1+K_{\rm NN}}}cK_{\rm NN}c(1+K_{\rm NN})X_1\sigma X_2\frac B{1+K_{\rm DD}}X_3\bar\sigma X_4 (1+K_{\rm NN})c\frac1{\sqrt{1+K_{\rm NN}}}\0\\
 &+\frac1{\sqrt{1+K_{\rm NN}}}c(1+K_{\rm NN})cX_1(K_{\rm ND}\sigma-\sigma K_{\rm ND})X_2 \frac B{1+K_{\rm DD}}X_3\bar\sigma X_4 (1+K_{\rm NN})c\frac1{\sqrt{1+K_{\rm NN}}}\0\\
 &+\frac1{\sqrt{1+K_{\rm NN}}}c(1+K_{\rm NN})X_1\sigma X_2 \frac {K_{\rm DD}}{1+K_{\rm DD}}X_3\bar\sigma X_4 (1+K_{\rm NN})c\frac1{\sqrt{1+K_{\rm NN}}}\0\\
 &-\frac1{\sqrt{1+K_{\rm NN}}}c(1+K_{\rm NN})X_1\sigma X_2\frac B{1+K_{\rm DD}}X_3(K_{\rm DN}\bar\sigma-K_{\rm DN}\bar\sigma)X_4 c(1+K_{\rm NN})c\frac1{\sqrt{1+K_{\rm NN}}}\0\\
 &-\frac1{\sqrt{1+K_{\rm NN}}}c(1+K_{\rm NN})X_1\sigma X_3\frac B{1+K_{\rm DD}}X_3\bar\sigma X_4(1+K_{\rm NN})cK_{\rm NN}c\frac1{\sqrt{1+K_{\rm NN}}},\0
 \end{align}
 where we have used the assumption that the $X_is$ are pure matter operators. Now with the help of \eqref{KXXK}, we can convert the $K_{ij}$ to $K_{ii}$  to obtain
 \begin{align}
 Q_{\rm NN}\Phi_0&=-\frac1{\sqrt{1+K_{\rm NN}}}cK_{\rm NN}c(1+K_{\rm NN})X_1\sigma X_2\frac B{1+K_{\rm DD}}X_3\bar\sigma X_4 (1+K_{\rm NN})c\frac1{\sqrt{1+K_{\rm NN}}}\0\\
 &+\frac1{\sqrt{1+K_{\rm NN}}}c(1+K_{\rm NN})cK_{\rm NN}X_1\sigma X_2 \frac B{1+K_{\rm DD}}X_3\bar\sigma X_4 (1+K_{\rm NN})c\frac1{\sqrt{1+K_{\rm NN}}}\0\\
 &-\frac1{\sqrt{1+K_{\rm NN}}}c(1+K_{\rm NN})cX_1\sigma X_2 K_{\rm DD}\frac B{1+K_{\rm DD}}X_3\bar\sigma X_4 (1+K_{\rm NN})c\frac1{\sqrt{1+K_{\rm NN}}}\0\\
 &+\frac1{\sqrt{1+K_{\rm NN}}}c(1+K_{\rm NN})X_1\sigma X_2 \frac {K_{\rm DD}}{1+K_{\rm DD}}X_3\bar\sigma X_4 (1+K_{\rm NN})c\frac1{\sqrt{1+K_{\rm NN}}}\0\\
 &-\frac1{\sqrt{1+K_{\rm NN}}}c(1+K_{\rm NN})X_1\sigma X_2\frac B{1+K_{\rm DD}}K_{\rm DD}X_3\bar\sigma X_4 c(1+K_{\rm NN})c\frac1{\sqrt{1+K_{\rm NN}}}\0\\
 &+\frac1{\sqrt{1+K_{\rm NN}}}c(1+K_{\rm NN})X_1\sigma X_2\frac B{1+K_{\rm DD}}X_3\bar\sigma X_4 K_{\rm NN}c(1+K_{\rm NN})c\frac1{\sqrt{1+K_{\rm NN}}}\0\\
 &-\frac1{\sqrt{1+K_{\rm NN}}}c(1+K_{\rm NN})X_1\sigma X_3\frac B{1+K_{\rm DD}}X_3\bar\sigma X_4(1+K_{\rm NN})cK_{\rm NN}c\frac1{\sqrt{1+K_{\rm NN}}}.\0
 \end{align}
 After some simplification this gives
 \begin{align}\label{QPhi02}
 Q_{\rm NN}\Phi_0&=-\frac1{\sqrt{1+K_{\rm NN}}}cK_{\rm NN}cX_1\sigma X_2\frac B{1+K_{\rm DD}}X_3\bar\sigma X_4 (1+K_{\rm NN})c\frac1{\sqrt{1+K_{\rm NN}}}\0\\
 &-\frac1{\sqrt{1+K_{\rm NN}}}cK_{\rm NN}cX_1\sigma X_2 K_{\rm DD}\frac {B}{1+K_{\rm DD}}X_3\bar\sigma X_4 (1+K_{\rm NN})c\frac1{\sqrt{1+K_{\rm NN}}}\0\\
 &+\frac1{\sqrt{1+K_{\rm NN}}}c(1+K_{\rm NN})X_1\sigma X_2 \frac {K_{\rm DD}}{1+K_{\rm DD}}X_3\bar\sigma X_4 (1+K_{\rm NN})c\frac1{\sqrt{1+K_{\rm NN}}}\0\\
 &-\frac1{\sqrt{1+K_{\rm NN}}}c(1+K_{\rm NN})X_1\sigma X_2 \frac {B}{1+K_{\rm DD}}{K_{\rm DD}}X_3\bar\sigma X_4 cK_{\rm NN}c\frac1{\sqrt{1+K_{\rm NN}}}\0\\
 &-\frac1{\sqrt{1+K_{\rm NN}}}c(1+K_{\rm NN})X_1\sigma X_2\frac B{1+K_{\rm DD}}X_3\bar\sigma X_4 cK_{\rm NN}c\frac1{\sqrt{1+K_{\rm NN}}}.
 \end{align}

 Similarly, using again the assumption that the $Xs$ are pure matter and also employing the OPE in \eqref{OPE3} we can write
 \begin{align}\label{Phi0Psi0}
 \{\Psi_0,\Phi_0\}&=\frac1{\sqrt{1+K_{\rm NN}}}cK_{\rm NN}cX_1\sigma X_2\frac B{1+K_{\rm DD}}X_3\bar\sigma X_4 (1+K_{\rm NN})c\frac1{\sqrt{1+K_{\rm NN}}}\0\\
 &+\frac1{\sqrt{1+K_{\rm NN}}}c(1+K_{\rm NN})X_1\sigma X_2\frac B{1+K_{\rm DD}}X_3\bar\sigma X_4 cK_{\rm NN}c\frac1{\sqrt{1+K_{\rm NN}}},
 \end{align}
 \begin{align}\label{Phi0^2}
 \Phi_0^2&=\frac1{\sqrt{1+K_{\rm NN}}}c(1+K_{\rm NN})X_1\sigma X_2Bc\frac 1{1+K_{\rm DD}}X_3\bar\sigma X_4 (1+K_{\rm NN})c\frac1{\sqrt{1+K_{\rm NN}}}\0\\
 &-\frac1{\sqrt{1+K_{\rm NN}}}c(1+K_{\rm NN})X_1\sigma X_2\frac 1{1+K_{\rm DD}}BcX_3\bar\sigma X_4 c(1+K_{\rm NN})\frac1{\sqrt{1+K_{\rm NN}}}.
 \end{align}
 Substituting the results in equations \eqref{Phi02}, \eqref{Phi0Psi0}, and \eqref{Phi0^2}, into the equation of motion of \eqref{TVEoM} and simplifying, we see that it is satisfied.

Since the equation of motion is satisfied, the energy is proportional to ${\rm Tr}[\Phi_0^3]$. Next, we calculate this trace. Lets replace $K_{\rm NN}$ by $K$ and $K_{\rm DD}$ by ${\cal K}$.
 \begin{align}
 {\rm Tr}\left[\Phi^3\right]=-&{\rm Tr}\Big[{c(1+K)}X_1\sigma X_2\frac1{1+{\cal K}}X_3\bar\sigma X_4(1+K)Bc\frac1{1+K}
 {c(1+K)}X_1\sigma X_2\frac1{1+{\cal K}}\0\\
 &\times X_3 \bar\sigma X_4(1+K)Bc\frac1{1+K}{c(1+K)}X_1\sigma X_2\frac1{1+{\cal K}}X_3 \bar\sigma X_4(1+K)Bc\frac1{1+K}\Big]\0
 \end{align}
 If we replace $c(1+K)=(1+K)c-\partial c$ we obtain,
 \begin{align}
 {\rm Tr}\left[\Phi^3\right]=&{\rm Tr}\Big[{\partial c}X_1\sigma X_2\frac1{1+{\cal K}}X_3 \bar\sigma X_4(1+K)Bc\frac1{1+K}
 {\partial c}X_1\sigma X_2\frac1{1+{\cal K}}\0\\
 &\times X_3 \bar\sigma X_4(1+K)Bc\frac1{1+K}{\partial c}X_1\sigma X_2\frac1{1+{\cal K}}X_3 \bar\sigma X_4(1+K)Bc\frac1{1+K}\Big],
 \end{align}
 where we have used $c^2=0$. Using $Bc+cB=1$, repeatedly, this can be further simplified as
 \begin{align}
 {\rm Tr}\left[\Phi^3\right]={\rm Tr}&\Big[X_1\sigma X_2{\partial c}\frac B{1+{\cal K}}X_3 \bar\sigma X_4
 X_1\sigma X_2{\partial c}\frac1{1+{\cal K}}X_3 \bar\sigma X_4X_1\sigma X_2{\partial c}\frac1{1+{\cal K}}\0\\
 &\times X_3 \bar\sigma X_4{(1+K)c}\frac1{1+K}\Big],
 \end{align}
 Recalling that $X_3 \bar\sigma X_4X_1\sigma X_2=1$ we obtain
 \begin{align}
 {\rm Tr}\left[\Phi^3\right]={\rm Tr}\left[\frac B{1+{\cal K}}
 {\partial c}\frac1{1+{\cal K}}{\partial c}\frac1{1+{\cal K}}X_3\bar\sigma X_4{(1+K)c}\frac1{1+K}X_1\sigma X_2{\partial c}\right],
 \end{align}
 where we have also applied cyclic property of the trace.
 Replacing $(1+K)c=c(1+K)+\partial c$ we obtain
 \begin{align}
 {\rm Tr}&\left[\Phi^3\right]={\rm Tr}\left[\frac B{1+{\cal K}}
 {\partial c}\frac1{1+{\cal K}}{\partial c}\frac1{1+{\cal K}}c{\partial c}\right]\0\\
 &+{\rm Tr}\left[\frac B{1+{\cal K}}
 {\partial c}\frac1{1+{\cal K}}{\partial c}\frac1{1+{\cal K}}X_3\bar\sigma X_4\partial c\frac1{1+K}X_1\sigma X_2{\partial c}\right],\nonumber\\
 &=\int_0^\infty dt_1dt_2dt_3e^{-(t_1+t_2+t_3)}\big\langle {\cal B}\partial c(0)\partial c(t_1)c\partial c(t_1+t_2)\big\rangle_{t_1+t_2+t_3}\Big(\langle 0|0\rangle_{\rm matter}\Big)\nonumber\\
 &+\int_0^\infty dt_1dt_2dt_3dt_4e^{-(t_1+t_2+t_3+t_4)}\big\langle {\cal B}\partial c(0)\partial c(t_1)\partial c(t_1+t_2)\partial c(t_1+t_2+t_3)\big\rangle_{t_1+t_2+t_3+t_4}\nonumber\\
 &~~~~~~~~~~~~~~~~~~~~~~~~~\times\big\langle X_3\bar\sigma X_4(t_1+t_2)X_1\sigma X_2(t_1+t_2+t_3)\big\rangle_{t_1+t_2+t_3+t_4}.
 \end{align}
where we have assumed that the wedge states defined in terms of ${\cal K}$ are the same as those of $K$. The ghost part of the second integrand is zero, while the first gives,
  \begin{align}
 &{\rm Tr}\left[\Phi_0^3\right]\0\\
 &=g_\ast\int_0^\infty dx_i\frac{e^{-(t_1+t_2+t_3)}}\pi\left[{\rm Sin}\left(\frac{2\pi (t_1+t_2)}{t_1+t_2+t_3}\right)-{\rm Sin}\left(\frac{2\pi t_1}{t_1+t_2+t_3}\right)-{\rm Sin}\left(\frac{2\pi t_2}{t_1+t_2+t_3}\right)\right].\0
 \end{align}
 Making the following usual change of variables
 \begin{align}
 x=\frac {t_1}{t_1+t_2+t_3},\quad y=\frac {t_2}{t_1+t_2+t_3},\quad T={t_1+t_2+t_3}
 \end{align}
 we can write
 \begin{align}
 &{\rm Tr}\left[\Phi_0^3\right]\0\\
&=-\frac{g_\ast}\pi\int_0^\infty dTT^2e^{-T}\int_0^1dx\int_0^{1-x}dy\left[{\rm Sin}\left(2\pi x\right)+{\rm Sin}\left(2\pi y\right)-{\rm Sin}\left(2\pi(x+y)\right)\right]=-\frac{3g_\ast}{\pi^2}
 \end{align}
 Therefore,
 \begin{align}
 E=-\frac16{\rm Tr}\left[\Phi_0^3\right]&=\frac{g_\ast}{2\pi^2}.
 \end{align}


\begin{thebibliography}{99}

\bibitem{Schnabl05}  M.~Schnabl,
 {\it Analytic solution for tachyon condensation in open string field theory,}
  Adv.\ Theor.\ Math.\ Phys.\  {\bf 10}, 433 (2006)
  [hep-th/0511286].

\bibitem{Ellwood}  I.~Ellwood  and M.~Schanabl,
    \textit {Proof of vanishing cohomology at the tachyon vacuum,}
  JHEP \textbf {0702} (2007) 096.


\bibitem{W}
  E.~Witten,
  {\it Noncommutative Geometry And String Field Theory,}
  Nucl.\ Phys.\  B {\bf 268} (1986) 253.


\bibitem{Okawa:2006vm}
  Y.~Okawa,
 {\it Comments on Schnabl's analytic solution for tachyon condensation in
  Witten's open string field theory,}
  JHEP {\bf 0604} (2006) 055
  [arXiv:hep-th/0603159].
\bibitem{ES}
  T.~Erler and M.~Schnabl,
  ``A Simple Analytic Solution for Tachyon Condensation,''
  JHEP {\bf 0910}, 066 (2009)
  [arXiv:0906.0979 [hep-th]].
\bibitem{KORZ}
  M.~Kiermaier, Y.~Okawa, L.~Rastelli and B.~Zwiebach,
 {\it Analytic solutions for marginal deformations in open string field theory,}
  arXiv:hep-th/0701249.

\bibitem{Okawa3}
  Y.~Okawa,
{\it Real analytic solutions for marginal deformations in open
superstring field theory,}
  arXiv:0704.3612 [hep-th].

\bibitem{Schnabl:2007az}
  M.~Schnabl,
 {\it Comments on marginal deformations in open string field theory,}
  arXiv:hep-th/0701248.




\bibitem{Fuchs3}
  E.~Fuchs, M.~Kroyter and R.~Potting,
  {\it Marginal deformations in string field theory,}
   arXiv:0704.2222 [hep-th].

\bibitem{Lee:2007ns}
  B.~H.~Lee, C.~Park and D.~D.~Tolla,
  {\it  Marginal Deformations as Lower Dimensional D-brane Solutions in Open
String
  Field theory,}
  arXiv:0710.1342 [hep-th].



\bibitem{Kiermaier:2007vu}
  M.~Kiermaier and Y.~Okawa,
  {\it Exact marginality in open string field theory: a general framework,}
  arXiv:0707.4472 [hep-th].


\bibitem{KOS} M.~Kiermaier, Y.~Okawa and P.~Soler,
{\it Solutions from boundary condition changing operators in open string field
theory,}
  JHEP {\bf 1103} (2011) 122
  [arXiv:1009.6185 [hep-th]].
\bibitem{Maccaferri:2014cpa}
  C.~Maccaferri,
  {\it A simple solution for marginal deformations in open string field theory,}
  JHEP {\bf 1405}, 004 (2014)
  [arXiv:1402.3546 [hep-th]].
	
	
\bibitem{Sen}
  A.~Sen,
 {\it Universality of the tachyon potential,}
  JHEP {\bf 9912}, 027 (1999)
  [arXiv:hep-th/9911116]


\bibitem{BMT} L.~Bonora, C.~Maccaferri and D.~D.~Tolla,
 {\it Relevant Deformations in Open String Field Theory: a Simple Solution for
  Lumps,} JHEP 1111:107,2011.   arXiv:1009.4158 [hep-th].

\bibitem{BGT1}
  L.~Bonora, S.~Giaccari and D.~D.~Tolla,
{\it The energy of the analytic lump solution in SFT,}
JHEP 08(2011)158. ArXiv:1105.5926 [hep-th]

\bibitem{BGT2}
  L.~Bonora, S.~Giaccari and D.~D.~Tolla,
{\it Analytic solutions for Dp branes in SFT,} JHEP12(2011)033.
 arXiv:1106.3914 [hep-th]

\bibitem{EM}
  T.~Erler and C.~Maccaferri,
{\it String Field Theory Solution for Any Open String Background,}
  JHEP {\bf 1410} (2014) 29
  [arXiv:1406.3021 [hep-th]].

  \bibitem{Fuchs} E.~Fuchs and M.~Kroyter,
{\it Analytical Solutions of Open String Field Theory,}
  Phys.\ Rept.\  {\bf 502} (2011) 89
  [arXiv:0807.4722 [hep-th]].

\bibitem{Schnabl}, M.~Schnabl,
{\it Algebraic solutions in Open String Field Theory - A Lightning Review},
[arXiv:1004.4858 [hep-th]].

\bibitem{Okawa} Y.~Okawa
{\it Analytic methods in open string field theory},
Prog.Theor.Phys. 128 (2012) 1001-1060.



\bibitem{Bonora} L.~Bonora
{\it String Field Theory: a short introduction}, PoS, ICMP2013 (2014) 001.

 \bibitem{Erler:2012qn} T.~Erler and C.~Maccaferri
{\it Connecting Solutions in Open String Field Theory with Singular Gauge
Transformations}
JHEP04(2012)107

 \bibitem{Cardy} L.~ Cardy, {\it Conformal invariance and surface critical
behavior}, Nucl. Phys. B {\bf 240} (1984) 514;
{\it Effect of boundary conditions on the operator content of two-dimensional
conformally invariant theories},
Nucl. Phys. B {\bf 275} (1986) 200;
{\it Boundary conditions, fusion rules and the Verlinde formula}, Nucl. Phys. B
{\bf 324} (1989) 581.

\bibitem{Zamolo} Al.~B.~ Zamolodchikov,
{\it Conformal scalar field on the hyperelliptic curve and critical
Ashkin-Teller multipoint correlation functions},
Nucl. Phys. B {\bf 285}(1987) 481.

\bibitem{Hashimoto} A.~ Hashimoto, {\it Dynamics of Dirichlet-Neumann open
strings on D-branes},
Nucl. Phys. B {\bf 496} (1997) 243, [hep-th/9608127].

\bibitem{GNS} E.~ Gava, K.~S.~ Narain, M.~H.~ Sarmadi,
{\it On the bound states of p- and (p+2)-branes}, Nucl. Phys. B {\bf 504} (1997)
214 [hep-th/9704006].

\bibitem{Frohlich}   J.~Frohlich, O.~Grandjean, A.~Recknagel and V.~Schomerus,
  {\it Fundamental strings in Dp - Dq brane systems,}   Nucl.\ Phys.\ B {\bf
583}, 381 (2000)
  [hep-th/9912079].

\bibitem{Dixon} L.~Dixon, D.~Friedan, E.~Martinec and S.~Shenker, {\it The
conformal field theory of orbifolds}
Nucl.Phys. B{\bf 282} (1987) 13.

 \bibitem{Corrigan} E.~Corrigan and D.~B.~Fairlie, {\it Off-shell staes in dual
resonance theory},
Nucl. Phys. B {\bf 91}(1975) 527.

\bibitem{Olive}
D.~Olive and J.~Scherk, {\it Towards satisfactory scattering
amplitudes for dual fermions},
Nucl. Phys. B {\bf 64}(1973) 334.
J~.~H.~Schwarz {\it Off-mass-shell dual amplitudes without ghosts},
Nucl. Phys. B {\bf 65}(1973) 131.
D.~Olive, P.Goddard, R.~A.~Smith and D.~J.~Olive, {\it Evaluation of the
scattering amplitude for four dual fermions}
Nucl. Phys. B {\bf 67}(1973) 477
J~.~H.~Schwarz, {\it Dual quark-gluon theory with dynamical color},
Nucl. Phys. B {\bf 68}(1974) 221.
E.~F.~Corrigan {\it The scattering amplitude for four dual fermions},
Nucl. Phys. B {\bf 69}(1974) 325.
J~.~H.~Schwarz and C.~C.~Wu {\it Off-shell dual amplitudes. II},
Nucl. Phys. B {\bf 72}(1974) 397.
M.~B.~Green {\it Locality and currents for the dual string},
Nucl. Phys. B {\bf 103}(1976) 333.











\bibitem{numlump} J.~A.~Harvey and P.~Kraus,
{\it D-Branes as unstable lumps in bosonic open string field theory,}
  JHEP {\bf 0004} (2000) 012, [hep-th/0002117].

R.~de Mello Koch, A.~Jevicki, M.~Mihailescu and R.~Tatar,
 {\it Lumps and p-branes in open string field theory,}
  Phys.\ Lett.\ B {\bf 482} (2000) 249, [hep-th/0003031].

N.~Moeller, A.~Sen and B.~Zwiebach,
{\it D-branes as tachyon lumps in string field theory,}
JHEP {\bf 0008} (2000) 039, [hep-th/0005036].

R.~de Mello Koch and J.~P.~Rodrigues,
{\it Lumps in level truncated open string field theory,}
Phys.\ Lett.\ B {\bf 495} (2000) 237, [hep-th/0008053].


\bibitem{RSZ} L.~Rastelli, A.~Sen and B.~Zwiebach,
{\it Classical solutions in string field theory around the tachyon vacuum,}
  Adv.\ Theor.\ Math.\ Phys.\  {\bf 5} (2002) 393, [hep-th/0102112].


\bibitem{Kutasov}
  D.~Kutasov, M.~Marino and G.~W.~Moore,
\textit {Some exact results on tachyon condensation in string field
theory,}   JHEP \textbf {0010} (2000) 045

\bibitem{Witten}
  E.~Witten,
 \textit {Some computations in background independent off-shell string theory,}
  Phys.\ Rev.\  D \textbf {47} (1993) 3405






\bibitem{BGT3}
  L.~Bonora, S.~Giaccari and D.~D.~Tolla,
{\it Lump solutions in SFT. Complements}
[arXiv:1109.4336, hep-th].


\bibitem{Kishimoto:2009nd}
  I.~Kishimoto and T.~Takahashi,
{\it Vacuum structure around identity based solutions,}
  Prog.\ Theor.\ Phys.\  {\bf 122}, 385 (2009)
  [arXiv:0904.1095 [hep-th]].
\bibitem{Zeze:2010jv}
  S.~Zeze,
  {\it Tachyon potential in KBc subalgebra,}
  Prog.\ Theor.\ Phys.\  {\bf 124}, 567 (2010)
  [arXiv:1004.4351 [hep-th]].

\bibitem{Zeze:2010sr}
  S.~Zeze,
  {\it Regularization of identity based solution in string field theory},
  arXiv:1008.1104 [hep-th].


\bibitem{Arroyo:2010sy} E.~A.~Arroyo,
  {\it Comments on regularization of identity based solutions in string field
  theory,}
  arXiv:1009.0198 [hep-th].



\bibitem{BG} L.~Bonora, S.~Giaccari
{\it Generalized states in SFT},
EPJC \textbf{73} (2013) 2644.






\end{thebibliography}


\end{document}